\documentclass[runningheads,a4paper]{llncs}

\usepackage{amssymb}
\usepackage{graphicx}

\usepackage{subfigure}
\usepackage{floatflt}

\usepackage{multirow}

\usepackage{color,soul}

\usepackage{listings}

\usepackage{tikz}
\usepackage{tikz-qtree}

\usepackage{csquotes} 

\usepackage{threeparttable}

\usepackage{url}
\urldef{\mailsa}\path|{todo}@xxx.com|
\newcommand{\keywords}[1]{\par\addvspace\baselineskip
	\noindent\keywordname\enspace\ignorespaces#1}

\usepackage{makeidx}
\makeindex

\ifnum\pdfoutput>0
\usepackage[
bookmarks=false,
breaklinks=true,
colorlinks=true,
linkcolor=black,
citecolor=black,
urlcolor=black,
pdfpagelayout=SinglePage
]{hyperref}
\usepackage[all]{hypcap}
\else
\usepackage{hyperref}
\fi

\usepackage[capitalise]{cleveref}
\crefname{section}{Sect.}{Sect.}
\Crefname{section}{Section}{Sections}
\crefname{figure}{Fig.}{Fig.}
\Crefname{figure}{Figure}{Figures}
\crefname{table}{Tab.}{Tab.}
\Crefname{table}{Table}{Tables}

\usepackage{xspace}
\newcommand{\eg}{e.\,g.,\ }
\newcommand{\ie}{i.\,e.,\ }

\begin{document}

\mainmatter  

\title{Toward A Collection of Cloud Integration Patterns}

\titlerunning{Toward A Collection of Cloud Integration Patterns}

\author{Daniel Ritter}
\institute{SAP SE, Dietmar-Hopp-Allee 16, 69190 Walldorf, Germany\\\email{daniel.ritter@sap.com}}

%
\iftrue
\author{Daniel Ritter\inst{1} \and Stefanie Rinderle-Ma\inst{2}}

\institute{
	SAP SE, Dietmar-Hopp-Allee 16, 69190 Walldorf, Germany\\\email{daniel.ritter@sap.com}\and
	University of Vienna, W\"ahringerstra\ss{}e 29, 1090 Wien, Austria\\\email{stefanie.rinderle-ma@univie.ac.at}
}
\fi

\toctitle{Lecture Notes in Computer Science}
\tocauthor{Authors' Instructions}
\maketitle
\begin{abstract}
Cloud computing is one of the most exciting IT trends nowadays. It poses several challenges on application integration with respect to, for example, security.
In this work we collect and categorize several new integration patterns and pattern solutions with a focus on cloud integration requirements.
Their evidence and examples are based on extensive literature and system reviews.


\end{abstract}

\section{Introduction}


In this work we revisit the current collection of \emph{Enterprise Integration Patterns} (EIP) in the context of cloud integration. On the basis of a quantitative analysis of several cloud integration scenarios on a well-established platform (\ie \cite{sap-hci}) 
, we conducted further qualitative literature and system reviews to collect and categorize additional characteristics. These characteristics are verified by cross-referencing them between the quantitative and the qualitative studies, categorized and formulated as a list of patterns or pattern solutions.

\section{Cloud Integration Patterns}
In this section we collect and define new integration patterns from the cloud integration domain as extensions to the EIPs \cite{Hohpe:2003:EIP:940308}. 
The pattern descriptions are represented in the format in \cref{tab:template}.

The categories we consider in this work are storage in \cref{sec:storage-pattern}, messaging patterns like transformation and routing in \cref{sec:messaging-pattern}, security in \cref{sec:security-pattern}, exception handling \cref{sec:exception-patterns}, monitoring and operations in \cref{sec:monitoring-pattern}, as well as adapter and endpoint patterns in \cref{sec:adapter-pattern}.

\subsection{Storage Patterns}
\label{sec:storage-pattern}

\begin{figure}[tbh]
	\centering
	\includegraphics[width=0.9\columnwidth]{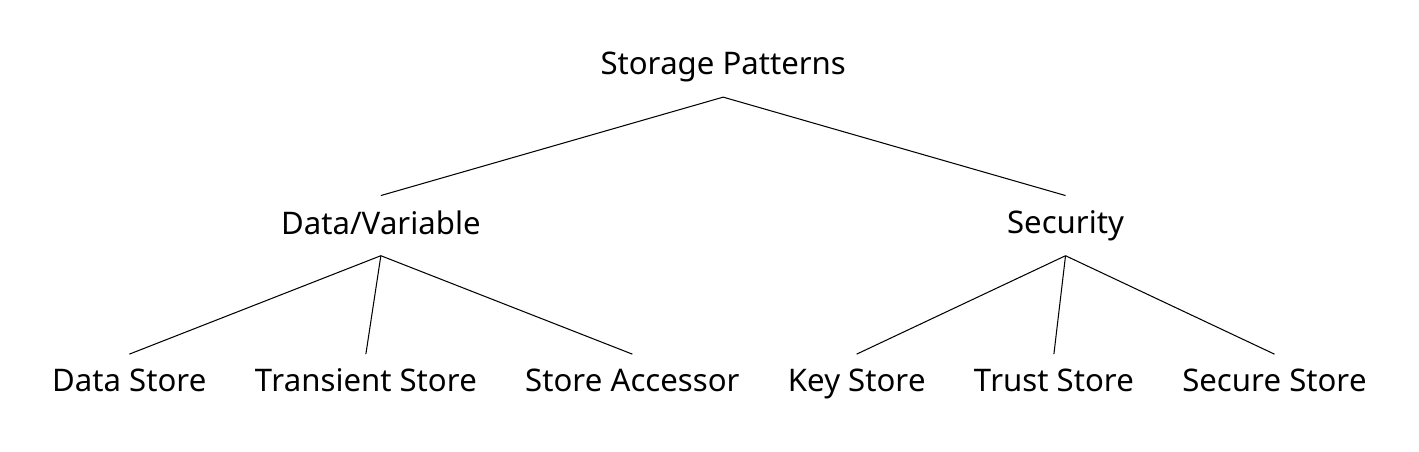}
	\caption{Storage Patterns.}
	\label{fig:storage-overview}
\end{figure}

In addition to the \emph{Message Store} \cite{Hohpe:2003:EIP:940308}, several vendors identified the need for further storage patterns as in \cref{fig:storage-overview}, \eg the storage of scheduler timings in \cref{sec:monitoring-pattern}, and persistent patterns like \emph{Aggregator} \cite{Hohpe:2003:EIP:940308} in a data store (cf. \cref{tab:st-patterns-4}).
During the processing, some vendors added capabilities to remember previous processing values, \eg variables, in a transient store \cref{tab:st-patterns-6}.
These stores can be accessed using a store accessor (cf. \cref{tab:st-patterns-5}).

\begin{table}
	\caption{Data Store \index{Data Store|textbf} \label{ceip:data-store}}
	\begin{tabular}{l*{2}{l}r}
		\hline
		Pattern Name          & Data Store \index{Data Store|textbf} \\
		\hline
		Intent                & \parbox[t]{0.9\columnwidth}{Store arbitrary content explicitly or implicitly from persistent patterns.} \\
		Driving Question      & \parbox[t]{0.9\columnwidth}{How can arbitrary data be stored persistent across instances for access by arbitrary patterns?} \\
		Solution              & \parbox[t]{0.9\columnwidth}{Provide a persistent store, which allows to store variables, value mapping, data of persistent patterns, \enquote{in-memory} storage capabilities.} \\
		Data Aspects          & \parbox[t]{0.9\columnwidth}{visibility: local (integration scenario), global (multiple integration scenarios or even solutions), multi-tenant, reliability: redundant, high-available, disaster recoverable; persistent} \\
		Variations            & \parbox[t]{0.9\columnwidth}{\eg relational column or row store, NoSQL store} \\
		Example               & \parbox[t]{0.9\columnwidth}{Persistent Scheduler, storage and archiving for legal regulations} \\
		Related Patterns      & \parbox[t]{0.9\columnwidth}{Message Store \cite{Hohpe:2003:EIP:940308}} \\
		Known Uses            & \parbox[t]{0.9\columnwidth}{partially covered by \enquote{Message Store} \cite{Ibsen:2010:CA:1965487}, \enquote{Flow reliability} \cite{apache-flume}, \enquote{StoreInKiteDataset} \cite{apache-nifi}, \enquote{DBStorage}, \enquote{Persist} \cite{sap-hci}} \\
	\end{tabular}
	\label{tab:st-patterns-4}
\end{table}

\begin{table}
	\caption{Transient Store \index{Transient Store|textbf} \label{ceip:transient-store}}
	\begin{tabular}{l*{2}{l}r}
		\hline
		Pattern Name          & Transient Store \index{Transient Store|textbf} \\
		\hline
		Intent                & \parbox[t]{0.9\columnwidth}{Remember arbitrary content explicitly within one integration scenario.} \\
		Driving Question      & \parbox[t]{0.9\columnwidth}{How to remember arbitrary data in one integration scenario context transiently?} \\
		Solution              & \parbox[t]{0.9\columnwidth}{Provide a transient store, which allows to remember variables, partial or complete messages in transient, \enquote{in-memory} storage capabilities.} \\
		Data Aspects          & \parbox[t]{0.9\columnwidth}{visibility: local to one integration scenario; transient; structured / unstructured} \\
		Variations            & \parbox[t]{0.9\columnwidth}{\eg relational column or row store, NoSQL store} \\
		Example               & \parbox[t]{0.9\columnwidth}{Variables} \\
		Related Patterns      & \parbox[t]{0.9\columnwidth}{Claim Check \cite{Hohpe:2003:EIP:940308}} \\
		Known Uses            & \parbox[t]{0.9\columnwidth}{partially covered by \enquote{Exchange Properties} \cite{Ibsen:2010:CA:1965487}, \enquote{Variables} \cite{sap-hci}} \\
	\end{tabular}
	\label{tab:st-patterns-6}
\end{table}

\begin{table}
	\caption{Store Accessor \index{Store Accessor|textbf} \label{ceip:store-accessor}}
	\begin{tabular}{l*{2}{l}r}
		\hline
		Pattern Name          & Store Accessor \index{Store Accessor|textbf} \\
		\hline
		Intent                & \parbox[t]{0.9\columnwidth}{Access the data store.} \\
		Driving Question      & \parbox[t]{0.9\columnwidth}{How to access the data store?} \\
		Data Aspects          & \parbox[t]{0.9\columnwidth}{transactions and data access depending on the data store, \eg query, read, write, delete} \\
		Example               & \parbox[t]{0.9\columnwidth}{Store timings of a Persistent Scheduler, aggregates of an Aggregator and persist (parts of) messages due to legal regulations for archiving.} \\
		Related Patterns      & \parbox[t]{0.9\columnwidth}{Data Store, Message Store \cite{Hohpe:2003:EIP:940308}, Claim Chek \cite{Hohpe:2003:EIP:940308}} \\
		Known Uses            & \parbox[t]{0.9\columnwidth}{\enquote{StoreInKiteDataset (Hadoop)} \cite{apache-nifi}, \enquote{DBStorage}, \enquote{Persist}, \enquote{Variables} \cite{sap-hci}} \\
	\end{tabular}
	\label{tab:st-patterns-5}
\end{table}


In addition, several security patterns from \cref{sec:security-pattern} require stored secure material like certificates, public/private keys in \cref{tab:st-patterns-1} or \cref{tab:st-patterns-2}, tokens and user/password in \cref{tab:st-patterns-3}.

\begin{table}
	\caption{Key Store \index{Key Store|textbf} \label{ceip:key-store}}
	\begin{tabular}{l*{2}{l}r}
		\hline
		Pattern Name          & Key Store \index{Key Store|textbf} \\
		\hline
		Intent                & \parbox[t]{0.9\columnwidth}{A store that contains private keys, and the certificates with their corresponding public keys.} \\
		Driving Question      & \parbox[t]{0.9\columnwidth}{How can keys and certificates be stored securely?} \\
		Solution              & \parbox[t]{0.9\columnwidth}{Provide a secured storage for security relevant key material, which is accessible from other patterns, but protected from remote access.} \\
		Data Aspects          & \parbox[t]{0.9\columnwidth}{certificates, persistent} \\
		Example               & \parbox[t]{0.9\columnwidth}{Set up and provide access to a Java Key Store from the integration content.} \\
		Related Patterns      & \parbox[t]{0.9\columnwidth}{Trust Store} \\
		Known Uses            & \parbox[t]{0.9\columnwidth}{\enquote{JKS} \cite{Ibsen:2010:CA:1965487}, \enquote{Key Store} \cite{apache-flume}, probably \cite{apache-nifi}, implicitly configurable in \cite{sap-hci}} \\
	\end{tabular}
	\label{tab:st-patterns-1}
\end{table}

\begin{table}
	\caption{Trust Store \index{Trust Store|textbf} \label{ceip:trust-store}}
	\begin{tabular}{l*{2}{l}r}
		\hline
		Pattern Name          & Trust Store \index{Trust Store|textbf} \\
		\hline
		Intent                & \parbox[t]{0.9\columnwidth}{A store that contains certificates from other parties that you expect to communicate with, or from Certificate Authorities that you trust to identify other parties.} \\
		Driving Question      & \parbox[t]{0.9\columnwidth}{How can certificates from other communication partners be stored securely?} \\
		Solution              & \parbox[t]{0.9\columnwidth}{Provide a secured storage for security relevant key material, which is accessible from other patterns. Probably limited access from respective parties is required to store their material.} \\
		Data Aspects          & \parbox[t]{0.9\columnwidth}{certificates, persistent} \\
		Related Patterns      & \parbox[t]{0.9\columnwidth}{Key Store} \\
		Known Uses            & \parbox[t]{0.9\columnwidth}{Trust Store \cite{apache-flume}} \\
	\end{tabular}
	\label{tab:st-patterns-2}
\end{table}

\begin{table}
	\caption{Secure Store \index{Secure Store|textbf} \label{ceip:plain-store}}
	\begin{tabular}{l*{2}{l}r}
		\hline
		Pattern Name          & Secure Store \index{Secure Store|textbf} \\
		\hline
		Intent                & \parbox[t]{0.9\columnwidth}{A store that contains users and their passwords or secure tokens.} \\
		Driving Question      & \parbox[t]{0.9\columnwidth}{How can users and their passwords or secure tokens be stored securely?} \\
		Solution              & \parbox[t]{0.9\columnwidth}{Secure and reliable storage for User/Password, Token, Expiring Token (cf. \cref{sec:security-pattern})} \\
		Data Aspects          & \parbox[t]{0.9\columnwidth}{user/password, token, persistent} \\
		Related Patterns      & \parbox[t]{0.9\columnwidth}{Key Store, Trust Store} \\
		Known Uses            & \parbox[t]{0.9\columnwidth}{\enquote{Secure Store} \cite{sap-hci}} \\
	\end{tabular}
	\label{tab:st-patterns-3}
\end{table}

\newpage
\subsection{Messaging Patterns}
\label{sec:messaging-pattern}

We combine the routing and transformation pattern categories from \cite{Hohpe:2003:EIP:940308} to general messaging patterns for the subsequent discussions.

\paragraph{Routing Patterns} For realizing, combined \emph{Scatter-Gather} \cite{Hohpe:2003:EIP:940308} (\eg for map-reduce like processing), the multicast is used \cref{tab:routing-patterns-1}.

Another new aspect is the communication between instances of several flows within one scenario or even between scenarios on the same platform (\ie no external call).
Therefore, a delegator can be used (cf. \cref{tab:routing-patterns-2}).
While in the asynchronous case, this can be realized by a messaging system endpoint, \eg JMS, many vendors offer special mechanisms to avoid an external call.

The iterative processing of tasks or explicit message re-deliveries requires a loop pattern (cf. \cref{tab:routing-patterns-3})

The structural combination of several channels to one (cf. \cref{tab:routing-patterns-4}) -- without merging messages -- is complementing the multicast pattern.

\begin{table}
	\caption{Multicast \index{Multicast|textbf} \label{ceip:multicast}}
	\begin{tabular}{l*{2}{l}r}
		\hline
		Pattern Name          & Multicast \index{Multicast|textbf} \\
		\hline
		Intent                & \parbox[t]{0.9\columnwidth}{Enable parallel message processing.} \\
		Driving Question      & \parbox[t]{0.9\columnwidth}{How to (statically) enable parallel message processing?} \\
		Solution              & \parbox[t]{0.9\columnwidth}{Send copies of a message to multiple receivers in parallel (statically)} \\
		Data Aspects          & \parbox[t]{0.9\columnwidth}{message creating, read-only, channel cardinality $1$:$n$, non-persistent} \\
		Variations            & \parbox[t]{0.9\columnwidth}{Parallel, Sequential Multicast} \\
		Related Patterns      & \parbox[t]{0.9\columnwidth}{Recipient List \cite{Hohpe:2003:EIP:940308}, Message Dispatcher, Splitter \cite{Hohpe:2003:EIP:940308}} \\
		Known Uses            & \parbox[t]{0.9\columnwidth}{\enquote{Multicast} \cite{Ibsen:2010:CA:1965487}, \enquote{Sequential/Parallel Multicast} \cite{sap-hci}} \\
	\end{tabular}
	\label{tab:routing-patterns-1}
\end{table}

\begin{table}
	\caption{Delegate \index{Delegate|textbf} \label{ceip:delegate}}
	\begin{tabular}{l*{2}{l}r}
		\hline
		Pattern Name          & Delegate \index{Delegate|textbf} \\
		\hline
		Intent                & \parbox[t]{0.9\columnwidth}{Exchange messages between several integration scenarios within the same integration system.} \\
		Driving Question      & \parbox[t]{0.9\columnwidth}{How to exchange messages between two or more integration scenarios locally?} \\
		Solution              & \parbox[t]{0.9\columnwidth}{Provide integration system local direct endpoints -- preferably using an optimized message exchange format and protocol.} \\
		Variations            & \parbox[t]{0.9\columnwidth}{Asynchronous message exchange via queues, synchronous message exchange via integration system local direct endpoints.} \\
		Example               & \parbox[t]{0.9\columnwidth}{Messaging System endpoint (\eg JMS, MQTT) for asynchronous messaging, or VM-local or platform local calls for synchrnonous messaging (potentially in an optimized format)} \\
		Related Patterns      & \parbox[t]{0.9\columnwidth}{Channel Adapter \cite{Hohpe:2003:EIP:940308}} \\
		Known Uses            & \parbox[t]{0.9\columnwidth}{\enquote{Direct-VM} endpoint \cite{Ibsen:2010:CA:1965487}, partially covered by \enquote{Process Call} \cite{sap-hci}} \\
	\end{tabular}
	\label{tab:routing-patterns-2}
\end{table}

\begin{table}
	\caption{Loop \index{Loop|textbf} \label{ceip:loop}}
	\begin{tabular}{l*{2}{l}r}
		\hline
		Pattern Name          & Loop \index{Loop|textbf} \\
		\hline
		Intent                & \parbox[t]{0.9\columnwidth}{Exchange messages between several integration scenarios within the same integration system.} \\
		Driving Question      & \parbox[t]{0.9\columnwidth}{How to exchange messages between two or more integration scenarios locally?} \\
		Solution              & \parbox[t]{0.9\columnwidth}{Provide integration system local direct endpoints -- preferably using an optimized message exchange format and protocol.} \\
		Variations            & \parbox[t]{0.9\columnwidth}{Sequential Loop, Interleaved Loop, Enclosing Loop.} \\
		Example               & \parbox[t]{0.9\columnwidth}{Iterative tasks like splitting, re-deliveries} \\
		Related Patterns      & \parbox[t]{0.9\columnwidth}{-} \\
		Known Uses            & \parbox[t]{0.9\columnwidth}{\enquote{Loop Activity} \cite{ibm}, \enquote{Looping} \cite{microsoft}} \\
	\end{tabular}
	\label{tab:routing-patterns-3}
\end{table}

\begin{table}
	\caption{Join Router \index{Join Router|textbf} \label{ceip:joinrouter}}
	\begin{tabular}{l*{2}{l}r}
		\hline
		Pattern Name          & Join Router \index{Join Router|textbf} \\
		\hline
		Intent                & \parbox[t]{0.9\columnwidth}{Combine several messages channels without merging the content.} \\
		Driving Question      & \parbox[t]{0.9\columnwidth}{How to combine several message channels without merging the content?} \\
		Solution              & \parbox[t]{0.9\columnwidth}{Provide a structural router, which receives messages of the same format from several incoming channels and forwards them to only one channel.} \\
		Related Patterns      & \parbox[t]{0.9\columnwidth}{opposite of multicast} \\
		Known Uses            & \parbox[t]{0.9\columnwidth}{\enquote{Direct-VM} \cite{Ibsen:2010:CA:1965487}, \enquote{Join} \cite{sap-hci}} \\
	\end{tabular}
	\label{tab:routing-patterns-4}
\end{table}

\begin{figure}[tbh]
	\centering
	\includegraphics[width=0.9\columnwidth]{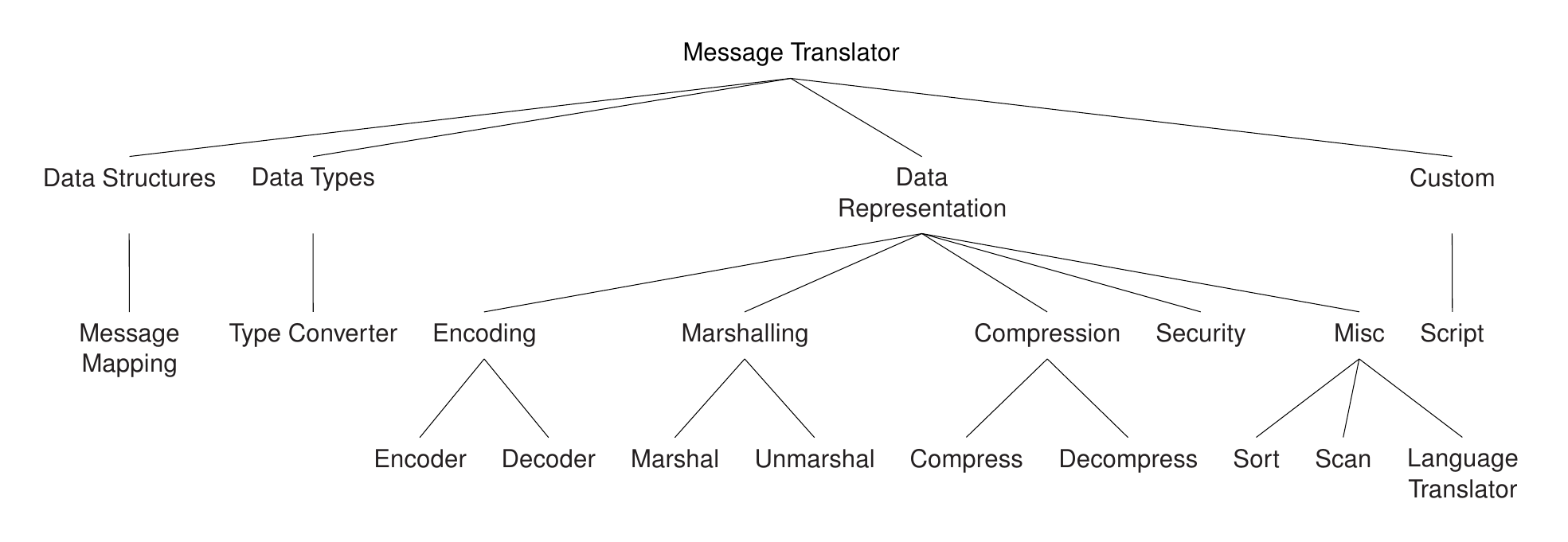}
	\caption{Message Translator Patterns}
	\label{fig:patterns-transformation}
\end{figure}

\paragraph{Multimedia / Transformation Patterns}
The message transformation patterns from \cite{Hohpe:2003:EIP:940308} are mainly \emph{Content Enricher}, \emph{Content Filter}, and \emph{Claim Check}. The \emph{Message Translator} itself is sketched around the OSI reference model in \cite{Hohpe:2003:EIP:940308}: data structures, data types, data representation and transport. Following these categories, we collected patterns as depicted in \cref{fig:patterns-transformation}. The security patterns enable message-level security and are further discussed in \cref{sec:security-pattern}. The custom processors are theoretically applicable to all categories.

The classical \emph{Message Translator} \cite{Hohpe:2003:EIP:940308} on data structure level is called message mapping and translates standardized data formats between applications. Typical implementations can be found, \eg as \enquote{Morphline Interceptor} in \cite{apache-flume} and the \enquote{Mapping} component in \cite{sap-hci}.

In practice, basic converters on data type level (cf. \cref{tab:tranform-patterns-1}) convert, \eg between different stream types, reader and writer types, byte arrays and string.

\begin{table}
	\caption{Type Converter \index{Type ConverterType Converter|textbf} \label{ceip:type_converter}}
	\begin{tabular}{l*{2}{l}r}
		\hline
		Pattern Name          & Type Converter \index{Type Converter|textbf} \\
		\hline
		Intent                & \parbox[t]{0.9\columnwidth}{Convert data types.} \\
		Driving Question      & \parbox[t]{0.9\columnwidth}{How to convert data types?} \\
		Solution              & \parbox[t]{0.9\columnwidth}{Provide special type converters for common types (domain specific).} \\
		Example               & \parbox[t]{0.9\columnwidth}{InputStream to OutputStream, Reader to Writer, ByteArray to String} \\
		Related Patterns      & \parbox[t]{0.9\columnwidth}{Custom Script} \\
		Known Uses            & \parbox[t]{0.9\columnwidth}{\enquote{Type Converter} \cite{Ibsen:2010:CA:1965487}, \cite{apache-nifi}, \enquote{Morphline Interceptor} \cite{apache-flume}, implicit in \cite{sap-hci}} \\
	\end{tabular}
	\label{tab:tranform-patterns-1}
\end{table}

On the data representation level these converters are complemented by transfer encoders to represent binary data in an ASCII string format (cf. \cref{tab:transform-patterns-2}) and decoders (cf. \cref{tab:transform-patterns-3}), as well as message protocol marshaller (cf. \cref{tab:transform-patterns-4}) and unmarshaller (cf. \cref{tab:transform-patterns-5}).

\begin{table}
	\caption{Encoder \index{Encoder|textbf} \label{ceip:encoder}}
	\begin{tabular}{l*{2}{l}r}
		\hline
		Pattern Name          & Encoder \index{Encoder|textbf} \\
		\hline
		Intent                & \parbox[t]{0.9\columnwidth}{Represent binary content textually.} \\
		Driving Question      & \parbox[t]{0.9\columnwidth}{How to represent binary content textually?} \\
		Solution              & \parbox[t]{0.9\columnwidth}{Provide standard and custom encoding capabilities for binary formats (domain-specific).} \\
		Data Aspects          & \parbox[t]{0.9\columnwidth}{Binary data encoding} \\
		Example               & \parbox[t]{0.9\columnwidth}{Base16, Base64 (ASCII, MIME content transfer encoding), Radix-64 (OpenPGP)} \\
		Related Patterns      & \parbox[t]{0.9\columnwidth}{Decoder} \\
		Known Uses            & \parbox[t]{0.9\columnwidth}{\enquote{Data Format} \cite{Ibsen:2010:CA:1965487}, \enquote{Morphline Interceptor} \cite{apache-flume}, \enquote{Base64EncodeContent} \cite{apache-nifi}, \enquote{Encoder} \cite{sap-hci}} \\
	\end{tabular}
	\label{tab:transform-patterns-2}
\end{table}

\begin{table}
	\caption{Decoder \index{Decoder|textbf} \label{ceip:decoder}}
	\begin{tabular}{l*{2}{l}r}
		\hline
		Pattern Name          & Decoder \index{Decoder|textbf} \\
		\hline
		Intent                & \parbox[t]{0.9\columnwidth}{Recover encoded binary content.} \\
		Driving Question      & \parbox[t]{0.9\columnwidth}{How to recover encoded binary content?} \\
		Solution              & \parbox[t]{0.9\columnwidth}{Provide standard and custom decoding capabilities for binary data (domain-specific).} \\
		Data Aspects          & \parbox[t]{0.9\columnwidth}{Encoding} \\
		Example               & \parbox[t]{0.9\columnwidth}{Base16, Base64 (ASCII, MIME content transfer encoding), Radix-64 (OpenPGP)} \\
		Related Patterns      & \parbox[t]{0.9\columnwidth}{Encoder} \\
		Known Uses            & \parbox[t]{0.9\columnwidth}{\enquote{Data Format} \cite{Ibsen:2010:CA:1965487}, \enquote{Morphline Interceptor} \cite{apache-flume}, \enquote{Base64EncodeContent} \cite{apache-nifi}, \enquote{Decoder} \cite{sap-hci}} \\
	\end{tabular}
	\label{tab:transform-patterns-3}
\end{table}

\begin{table}
	\caption{Marshaller \index{Marshaller|textbf} \label{ceip:marshaller}}
	\begin{tabular}{l*{2}{l}r}
		\hline
		Pattern Name          & Marshaller \index{Marshaller|textbf} \\
		\hline
		Intent                & \parbox[t]{0.9\columnwidth}{Convert data formats.} \\
		Driving Question      & \parbox[t]{0.9\columnwidth}{How to convert one data format to another?} \\
		Solution              & \parbox[t]{0.9\columnwidth}{Provide standard and custom marshalling capabilities for data formats (domain-specific).} \\
		Data Aspects          & \parbox[t]{0.9\columnwidth}{Encoding} \\
		Example               & \parbox[t]{0.9\columnwidth}{JSON, XML, POJO, SQL} \\
		Related Patterns      & \parbox[t]{0.9\columnwidth}{Unmarshaller} \\
		Known Uses            & \parbox[t]{0.9\columnwidth}{\enquote{Data Format} \cite{Ibsen:2010:CA:1965487}, \enquote{Morphline Interceptor} \cite{apache-flume}, \enquote{ConvertJSONToSQL} \cite{apache-nifi}, \enquote{JsonXMLConverter} \cite{sap-hci}} \\
	\end{tabular}
	\label{tab:transform-patterns-4}
\end{table}

\begin{table}
	\caption{Unmarshaller \index{Unmarshaller|textbf} \label{ceip:unmarshaller}}
	\begin{tabular}{l*{2}{l}r}
		\hline
		Pattern Name          & Unmarshaller \index{Unmarshaller|textbf} \\
		\hline
		Intent                & \parbox[t]{0.9\columnwidth}{Convert data formats.} \\
		Driving Question      & \parbox[t]{0.9\columnwidth}{How to convert one data format to another?} \\
		Solution              & \parbox[t]{0.9\columnwidth}{Provide standard and custom unmarshalling capabilities for data formats (domain-specific).} \\
		Data Aspects          & \parbox[t]{0.9\columnwidth}{Encoding} \\
		Example               & \parbox[t]{0.9\columnwidth}{JSON, XML, POJO, SQL} \\
		Related Patterns      & \parbox[t]{0.9\columnwidth}{Unmarshaller} \\
		Known Uses            & \parbox[t]{0.9\columnwidth}{\enquote{Data Format} \cite{Ibsen:2010:CA:1965487}, \enquote{Morphline Interceptor} \cite{apache-flume}, \enquote{ConvertJSONToSQL} \cite{apache-nifi}, \enquote{JsonXMLConverter} \cite{sap-hci}} \\
	\end{tabular}
	\label{tab:transform-patterns-5}
\end{table}

One way to reduce the amount of exchanged data is to compress (cf. \cref{tab:transform-patterns-5}) and later decompress the content (cf. \cref{tab:transform-patterns-6}).
The operations on binary, unstructured data is required, \eg for resizing an image for interoperability or compression (cf. \cref{tab:transform-patterns-108})
Furthermore, the extraction of metadata from arbitrary non-structured documents is required for multimedia scenarios (cf. \cref{tab:transform-patterns-107}).

\begin{table}
	\caption{Compress Content \index{Compress Content|textbf} \label{ceip:compress}}
	\begin{tabular}{l*{2}{l}r}
		\hline
		Pattern Name          & Compress Content \index{Compress Content|textbf} \\
		\hline
		Intent                & \parbox[t]{0.9\columnwidth}{Compress message content.} \\
		Driving Question      & \parbox[t]{0.9\columnwidth}{How to compress the message content?} \\
		Solution              & \parbox[t]{0.9\columnwidth}{Provide compression capabilities that compress (parts of) messages.} \\
		Data Aspects          & \parbox[t]{0.9\columnwidth}{Compression algorithm, type} \\
		Example               & \parbox[t]{0.9\columnwidth}{GZIP} \\
		Related Patterns      & \parbox[t]{0.9\columnwidth}{Decompress Content, Message Endpoint, Adapter} \\
		Known Uses            & \parbox[t]{0.9\columnwidth}{\enquote{Morphline Interceptor} \cite{apache-flume}, \enquote{Compress Content} \cite{apache-nifi}, \enquote{Custom Script} \cite{sap-hci}} \\
	\end{tabular}
	\label{tab:transform-patterns-5}
\end{table}

\begin{table}
	\caption{Decompress Content \index{Decompress Content|textbf} \label{ceip:decompress}}
	\begin{tabular}{l*{2}{l}r}
		\hline
		Pattern Name          & Decompress Content \index{Decompress Content|textbf} \\
		\hline
		Intent                & \parbox[t]{0.9\columnwidth}{Decompress message content.} \\
		Driving Question      & \parbox[t]{0.9\columnwidth}{How to decompress message content?} \\
		Solution              & \parbox[t]{0.9\columnwidth}{Provide decompression capabilities that decompress (parts of) messages.} \\
		Data Aspects          & \parbox[t]{0.9\columnwidth}{Decompression algorithm, type} \\
		Example               & \parbox[t]{0.9\columnwidth}{GZIP} \\
		Related Patterns      & \parbox[t]{0.9\columnwidth}{Compress Content, Message Endpoint, Adapter} \\
		Known Uses            & \parbox[t]{0.9\columnwidth}{\enquote{Morphline Interceptor} \cite{apache-flume}, \enquote{Compress Content} \cite{apache-nifi}, \enquote{Custom Script} \cite{sap-hci}} \\
	\end{tabular}
	\label{tab:transform-patterns-6}
\end{table}

\begin{table}
	\caption{Image Resizer \index{Image Resizer|textbf} \label{ceip:resizer}}
	\begin{tabular}{l*{2}{l}r}
		\hline
		Pattern Name          & Image Resizer \index{Image Resizer|textbf} \\
		\hline
		Intent                & \parbox[t]{0.9\columnwidth}{Resize Image.} \\
		Driving Question      & \parbox[t]{0.9\columnwidth}{How to resize an image?} \\
		Solution              & \parbox[t]{0.9\columnwidth}{Scale image using image processing capabilities.} \\
		Data Aspects          & \parbox[t]{0.9\columnwidth}{binary, image data, size information} \\
		Example               & \parbox[t]{0.9\columnwidth}{Resize image for interoperability with multimedia endpoint like Facebook or Twitter or compression.} \\
		Related Patterns      & \parbox[t]{0.9\columnwidth}{Compress Content} \\
		Known Uses            & \parbox[t]{0.9\columnwidth}{\enquote{Image Resizer} \cite{apache-nifi}} \\
	\end{tabular}
	\label{tab:transform-patterns-108}
\end{table}

\begin{table}
	\caption{Metadata Extractor \index{Metadata Extractor|textbf} \label{ceip:metaextractor}}
	\begin{tabular}{l*{2}{l}r}
		\hline
		Pattern Name          & Metadata Extractor \index{Metadata Extractor|textbf} \\
		\hline
		Intent                & \parbox[t]{0.9\columnwidth}{Extract multimedia meta data.} \\
		Driving Question      & \parbox[t]{0.9\columnwidth}{How to extract multimedia metadata?} \\
		Solution              & \parbox[t]{0.9\columnwidth}{Provide extraction capabilities that get metadata of multimedia data (\eg image, video, document).} \\
		Data Aspects          & \parbox[t]{0.9\columnwidth}{Metadata like geographic location, time} \\
		Example               & \parbox[t]{0.9\columnwidth}{Geo tag} \\
		Related Patterns      & \parbox[t]{0.9\columnwidth}{-} \\
		Known Uses            & \parbox[t]{0.9\columnwidth}{\enquote{Read MIME activity} \cite{ibm}, \enquote{ExtractImageMetadata}, \enquote{ExtractMediaMetadata} \cite{apache-nifi}} \\
	\end{tabular}
	\label{tab:transform-patterns-107}
\end{table}

Some special constructs that were not picked up in general are a special sort pattern (cf. \cref{tab:transform-patterns-8}), a scanner (cf. \cref{tab:transform-patterns-9}) and an actual language translator (cf. \cref{tab:transform-patterns-10}).

\begin{table}
	\caption{Content Sort \index{Content Sort|textbf} \label{ceip:sort}}
	\begin{tabular}{l*{2}{l}r}
		\hline
		Pattern Name          & Content Sort \index{Content Sort|textbf} \\
		\hline
		Intent                & \parbox[t]{0.9\columnwidth}{Sort message content.} \\
		Driving Question      & \parbox[t]{0.9\columnwidth}{How to sort the content of a message?} \\
		Solution              & \parbox[t]{0.9\columnwidth}{Provide configurable sort capabilities that access the message content.} \\
		Data Aspects          & \parbox[t]{0.9\columnwidth}{Algorithms, Comparators} \\
		Example               & \parbox[t]{0.9\columnwidth}{User default comparator in the programming language to sort alpha-numeric content.} \\
		Related Patterns      & \parbox[t]{0.9\columnwidth}{Custom Custom Script, Aggregator} \\
		Known Uses            & \parbox[t]{0.9\columnwidth}{\enquote{Sort} \cite{Ibsen:2010:CA:1965487}, \enquote{Morphline Interceptor} \cite{apache-flume}, \enquote{Custom Script} \cite{sap-hci}} \\
	\end{tabular}
	\label{tab:transform-patterns-8}
\end{table}

\begin{table}
	\caption{Find and Replace \index{Find and Replace|textbf} \label{ceip:scanner}}
	\begin{tabular}{l*{2}{l}r}
		\hline
		Pattern Name          & Find and Replace \index{Find and Replace|textbf} \\
		\hline
		Intent                & \parbox[t]{0.9\columnwidth}{Find and replace content.} \\
		Driving Question      & \parbox[t]{0.9\columnwidth}{How to find and replace content?} \\
		Solution              & \parbox[t]{0.9\columnwidth}{Scans the content of a message (for terms that are found in a user-supplied dictionary). If a term is matched, the UTF-8 encoded version of the term will be added to the message.} \\
		Data Aspects          & \parbox[t]{0.9\columnwidth}{optional dictionary} \\
		Example               & \parbox[t]{0.9\columnwidth}{Regex} \\
		Related Patterns      & \parbox[t]{0.9\columnwidth}{Custom Custom Script} \\
		Known Uses            & \parbox[t]{0.9\columnwidth}{\enquote{Scan Content} \cite{apache-nifi}, \enquote{Morphline Interceptor} \cite{apache-flume}, \enquote{Processor} \cite{Ibsen:2010:CA:1965487}, \enquote{Custom Script} \cite{sap-hci}} \\
	\end{tabular}
	\label{tab:transform-patterns-9}
\end{table}

\begin{table}
	\caption{Language Translator \index{Language Translator|textbf} \label{ceip:languagetranslator}}
	\begin{tabular}{l*{2}{l}r}
		\hline
		Pattern Name          & Language Translator \index{Language Translator|textbf} \\
		\hline
		Intent                & \parbox[t]{0.9\columnwidth}{Translate textual content from one language to another.} \\
		Driving Question      & \parbox[t]{0.9\columnwidth}{How to translate textual content from one language to another?} \\
		Solution              & \parbox[t]{0.9\columnwidth}{Provide standard interface access to natural language translation libraries or services.} \\
		Data Aspects          & \parbox[t]{0.9\columnwidth}{Service bindings, locale} \\
		Example               & \parbox[t]{0.9\columnwidth}{Yandex or Google translator API.} \\
		Related Patterns      & \parbox[t]{0.9\columnwidth}{Custom Custom Script} \\
		Known Uses            & \parbox[t]{0.9\columnwidth}{\enquote{Yandex Translator} \cite{apache-nifi}, \enquote{Morphline Interceptor} \cite{apache-flume}, \enquote{Processor} \cite{Ibsen:2010:CA:1965487}, \enquote{Custom Script} \cite{sap-hci}} \\
	\end{tabular}
	\label{tab:transform-patterns-10}
\end{table}

Besides the specialized translation patterns, language bindings for the execution of custom code are required for flexibility (cf. \cref{tab:transform-patterns-7}).

\begin{table}
	\caption{Custom Custom Script \index{Custom Custom Script|textbf} \label{ceip:script}}
	\begin{tabular}{l*{2}{l}r}
		\hline
		Pattern Name          & Custom Custom Script \index{Custom Custom Script|textbf} \\
		\hline
		Intent                & \parbox[t]{0.9\columnwidth}{Provide arbitrary body, header and attachment access and allow modifications.} \\
		Driving Question      & \parbox[t]{0.9\columnwidth}{How to execute custom code during message processing?} \\
		Solution              & \parbox[t]{0.9\columnwidth}{Embed freely programmable language support with access to the message channel and the messages.} \\
		Data Aspects          & \parbox[t]{0.9\columnwidth}{Service bindings, compiled or plain source code (interpreted)} \\
		Variations            & \parbox[t]{0.9\columnwidth}{controlled access by pre-defined, limiting expression language.} \\
		Example               & \parbox[t]{0.9\columnwidth}{Language support for Groovy or Java script with access to the message and other services of the integration system \eg Store Accessor.} \\
		Related Patterns      & \parbox[t]{0.9\columnwidth}{Content Modifier, Content Enricher} \\
		Known Uses            & \parbox[t]{0.9\columnwidth}{\enquote{Morphline Interceptor} \cite{apache-flume}, \enquote{Processor} \cite{Ibsen:2010:CA:1965487}, \enquote{Custom Script} \cite{sap-hci}} \\
	\end{tabular}
	\label{tab:transform-patterns-7}
\end{table}




\newpage
\subsection{Security Patterns}
\label{sec:security-pattern}

\begin{figure}[tbh]
	\centering
	\includegraphics[width=1.0\columnwidth]{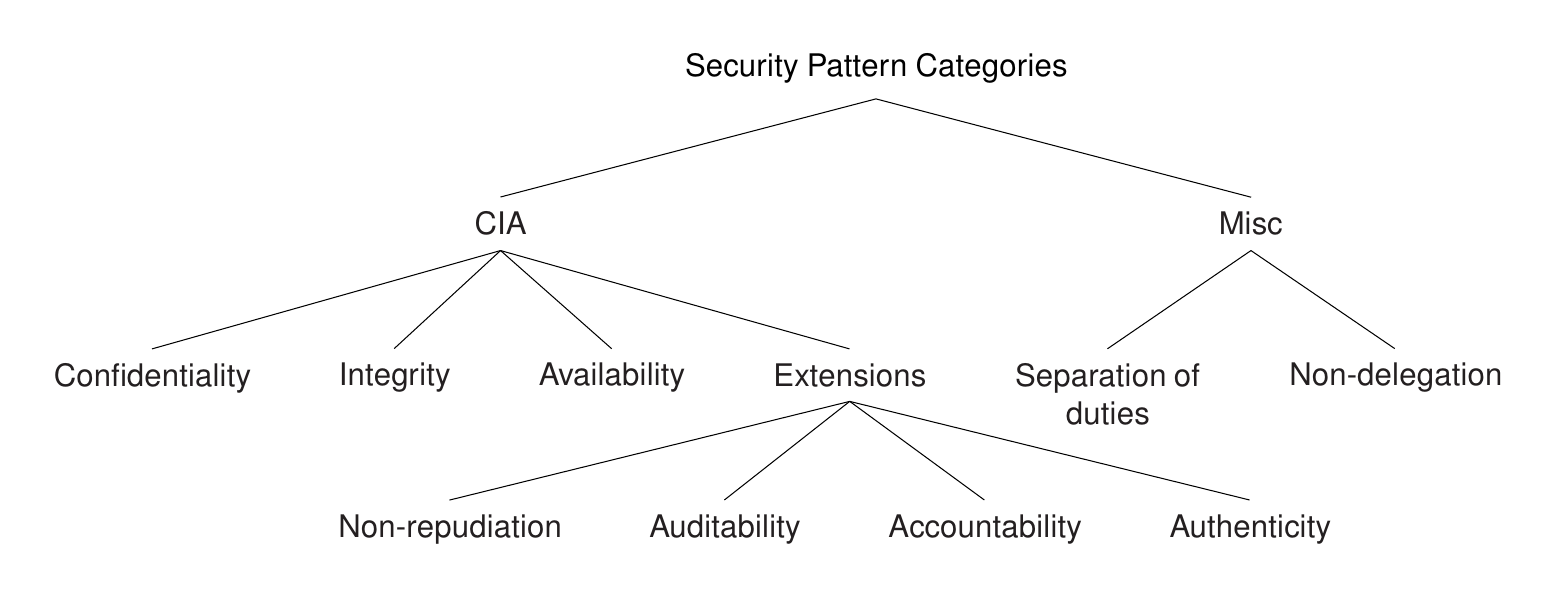}
	\caption{Security Categories grouped by classical CIA and extensions \cite{Whitman:2007:PIS:1557352}.}
	\label{fig:security_patterns}
\end{figure}

The collection of security patterns is grounded our system and literature study on security aspects in integration systems. 
Subsequently, the security patterns are listed and defined grouped by the security categories shown in \cref{fig:security_patterns}.

\begin{figure}[tbh]
	\centering
	\includegraphics[width=1.0\columnwidth]{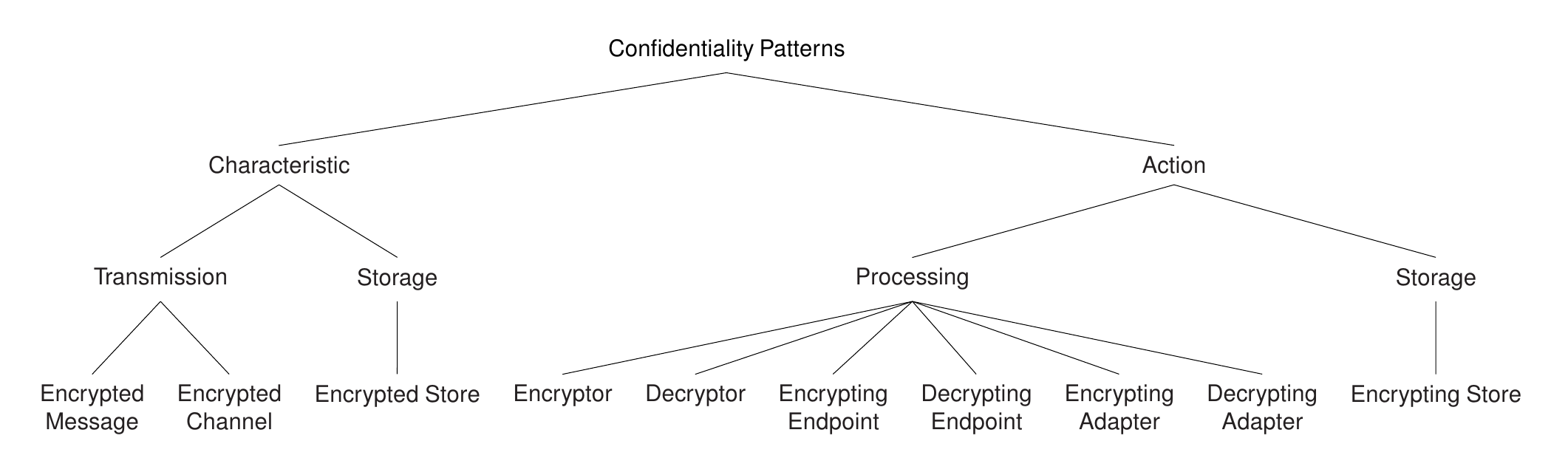}
	\caption{Confidentiality Patterns grouped by characteristics and actions 
		as well as information states: processing, transmission and storage \cite{mccumber1991information}.}
	\label{fig:security_confidentiality_patterns}
\end{figure}

\paragraph{Confidentiality Patterns} The confidentiality or message privacy patterns collected in \cref{fig:security_confidentiality_patterns} help to ensure that only authorized participants can access integration information like messages, channels, operations and storage.
The message \cref{ceip:encrypted-message}, the channel \cref{ceip:encrypted-channel} and the storage \cref{ceip:encrypted-store} can have the characteristic of having encrypted information (\eg encryption \cite{DBLP:conf/ifip8-1/LeitnerSRS13}).

The security-action patterns consist of pair-wise encrypting and decrypting operations (\cref{ceip:encryptor}, \cref{ceip:decryptor}), endpoints (\cref{ceip:encrypting-endpoint}, \cref{ceip:decrypting-endpoint}), adapters (\cref{ceip:encrypting-adapter}, \cref{ceip:decrypting-adapter}) and an encrypting store (\cref{ceip:encrypting-store}).

\begin{table}
	\caption{Encrypted Message \index{Encrypted Message|textbf} \label{ceip:encrypted-message}}
	\begin{tabular}{l*{2}{l}r}
		\hline
		Pattern Name          & Encrypted Message \index{Encrypted Message|textbf} \label{ceip:encrypted-message} \\
		\hline
		Intent                & \parbox[t]{0.9\columnwidth}{Rely on a confidential and private piece of information.} \\
		Driving Question      & \parbox[t]{0.9\columnwidth}{How can messages be sent confidential and with data-privacy?} \\
		Context               & \parbox[t]{0.9\columnwidth}{The confidentiality or data-privacy of a message is especially important, when the communication happens via a public network (\eg cloud integration).} \\
		Solution              & \parbox[t]{0.9\columnwidth}{Message level security; distinguish symmetric, asymmetric; text and categories (categories: channel, message cardinality, input/output, message generating, read/write access).} \\
		Data Aspects          & \parbox[t]{0.9\columnwidth}{text and categories (categories: channel, message cardinality, input/output, message generating, read/write access)} \\
		Result                & \parbox[t]{0.9\columnwidth}{The message is asymmetrically encrypted. The message can only be read by applying, \eg a \emph{Decryptor} pattern}. \\
		Example               & \parbox[t]{0.9\columnwidth}{PGP, PCKS7} \\
		Related Patterns      & \parbox[t]{0.9\columnwidth}{Message, Message Encryptor} \\
		Known Uses            & \parbox[t]{0.9\columnwidth}{\enquote{PGP Message} \cite{Ibsen:2010:CA:1965487}, \enquote{Encrypted Content} \cite{apache-nifi}, implicitly in \cite{sap-hci}} \\
	\end{tabular}
	\label{tab:sec-patterns-1}
\end{table}


\begin{table}
	\caption{Encrypted Channel \index{Encrypted Channel|textbf} \label{ceip:encrypted-channel}}
	\begin{tabular}{l*{2}{l}r}
		\hline
		Pattern Name          & Encrypted Channel \index{Encrypted Channel|textbf} \\
		\hline
		Intent                & \parbox[t]{0.9\columnwidth}{Exchange message over a secure channel.} \\
		Driving Question      & \parbox[t]{0.9\columnwidth}{How can an application send a confidential message such that only the actual receiver can process it?} \\
		Solution              & \parbox[t]{0.9\columnwidth}{Transport level security} \\
		Data Aspects          & \parbox[t]{0.9\columnwidth}{Encrypted transfer, certificates} \\
		Example               & \parbox[t]{0.9\columnwidth}{SSL/TLS, HTTPS} \\
		Related Patterns      & \parbox[t]{0.9\columnwidth}{Encrypted Message} \\
		Known Uses            & \parbox[t]{0.9\columnwidth}{\cite{apache-flume}, \cite{apache-nifi}, \cite{Ibsen:2010:CA:1965487}, \cite{sap-hci}} \\
	\end{tabular}
	\label{tab:sec-patterns-2}
\end{table}


\begin{table}
	\caption{Encrypted Store \index{Encrypted Store|textbf} \label{ceip:encrypted-store}}
	\begin{tabular}{l*{2}{l}r}
		\hline
		Pattern Name          & Encrypted Store \index{Encrypted Store|textbf} \\
		\hline
		Intent                & \parbox[t]{0.9\columnwidth}{Store messages confidential.} \\
		Driving Question      & \parbox[t]{0.9\columnwidth}{How to store messages confidential?} \\
		Solution              & \parbox[t]{0.9\columnwidth}{Message level security; TODO: stores messages encrypted} \\
		Data Aspects          & \parbox[t]{0.9\columnwidth}{Encrypted content, header and attachments} \\
		Example               & \parbox[t]{0.9\columnwidth}{PGP, PCKS7} \\
		Related Patterns      & \parbox[t]{0.9\columnwidth}{Message Encryptor, Encrypted Message} \\
		Known Uses            & \parbox[t]{0.9\columnwidth}{Configuration on \enquote{DBStorage} \cite{sap-hci}} \\
	\end{tabular}
	\label{tab:sec-patterns-3}
\end{table}

\begin{table}
	\caption{Message Encryptor \index{Message Encryptor|textbf} \label{ceip:encryptor}}
	\begin{tabular}{l*{2}{l}r}
		\hline
		Pattern Name          & Message Encryptor  \index{Message Encryptor|textbf} \\
		\hline
		Intent                & \parbox[t]{0.9\columnwidth}{When an application sends a confidential message to another participant, its content shall only be read by the receiver.} \\
		Driving Question      & \parbox[t]{0.9\columnwidth}{How can messages be sent confidential and with message-privacy as \emph{Encrypted Message}?} \\
		Solution              & \parbox[t]{0.9\columnwidth}{Provide capabilities to encrypt the content, headers and/or attachments of a message.} \\
		Data Aspects          & \parbox[t]{0.9\columnwidth}{in: message, out: encrypted message, key store, public key, non-message generating, content modifying} \\
		Example               & \parbox[t]{0.9\columnwidth}{PGP, PCKS7} \\
		Related Patterns      & \parbox[t]{0.9\columnwidth}{Encrypted Message, Message Decryptor} \\
		Known Uses            & \parbox[t]{0.9\columnwidth}{\enquote{Encrypt content} \cite{apache-nifi}, \enquote{Encryptor} \cite{sap-hci}} \\
	\end{tabular}
	\label{tab:sec-patterns-4}
\end{table}

\begin{table}
	\caption{Message Decryptor \index{Message Decryptor|textbf} \label{ceip:decryptor}}
	\begin{tabular}{l*{2}{l}r}
		\hline
		Pattern Name          & Message Decryptor \index{Message Decryptor|textbf} \label{ceip:decryptor} \\
		\hline
		Intent                & \parbox[t]{0.9\columnwidth}{When an application sends a confidential message to another participant, its content shall only be read by the receiver.} \\
		Driving Question      & \parbox[t]{0.9\columnwidth}{How can confidential messages \emph{Encrypted Message} be processed by the actual receiver?} \\
		Solution              & \parbox[t]{0.9\columnwidth}{Provide capabilities to decrypt the content, headers and/or attachments of a message.} \\
		Data Aspects          & \parbox[t]{0.9\columnwidth}{in: encrypted message, out: decrypted message, key store, private key, non-message generating, content modifying.} \\
		Example               & \parbox[t]{0.9\columnwidth}{PGP, PCKS7} \\
		Related Patterns      & \parbox[t]{0.9\columnwidth}{Message Encryptor} \\
		Known Uses            & \parbox[t]{0.9\columnwidth}{\enquote{Decrypt content} \cite{apache-nifi}, \enquote{Decryptor} \cite{sap-hci}} \\
	\end{tabular}
	\label{tab:sec-patterns-5}
\end{table}

\begin{table}
	\caption{Encrypting Endpoint \index{Encrypting Endpoint|textbf} \label{ceip:encrypting-endpoint}}
	\begin{tabular}{l*{2}{l}r}
		\hline
		Pattern Name          & Encrypting Endpoint \index{Encrypting Endpoint|textbf} \label{ceip:encrypting-endpoint} \\
		\hline
		Intent                & \parbox[t]{0.9\columnwidth}{When an application sends a confidential message to another participant, its content shall only be read by the receiver.} \\
		Driving Question      & \parbox[t]{0.9\columnwidth}{How can messages be sent confidential and with message-privacy as \emph{Encrypted Message}?} \\
		Solution              & \parbox[t]{0.9\columnwidth}{Provide capabilities to encrypt the content, headers and/or attachments of a message.} \\
		Data Aspects          & \parbox[t]{0.9\columnwidth}{key store, public key, non-message generating, content modifying} \\
		Example               & \parbox[t]{0.9\columnwidth}{PGP, PCKS7} \\
		Related Patterns      & \parbox[t]{0.9\columnwidth}{Message Encryptor} \\
		Known Uses            & \parbox[t]{0.9\columnwidth}{\cite{sap-hci}} \\
	\end{tabular}
	\label{tab:sec-patterns-6}
\end{table}

\begin{table} \label{todo}
	\caption{Decrypting Endpoint \index{Decrypting Endpoint|textbf} \label{ceip:decrypting-endpoint}} \label{todo}
	\begin{tabular}{l*{2}{l}r}
		\hline
		Pattern Name          & Decrypting Endpoint \index{Decrypting Endpoint|textbf} \label{ceip:decrypting-endpoint} \\
		\hline
		Intent                & \parbox[t]{0.9\columnwidth}{When an application sends a confidential message to another participant, its content shall only be read by the receiver.} \\
		Driving Question      & \parbox[t]{0.9\columnwidth}{How can confidential messages \emph{Encrypted Message} be processed by the actual receiver?} \\
		Solution              & \parbox[t]{0.9\columnwidth}{Provide capabilities to decrypt the content, headers and/or attachments of a message.} \\
		Data Aspects          & \parbox[t]{0.9\columnwidth}{key store, private key, non-message generating, content modifying} \\
		Example               & \parbox[t]{0.9\columnwidth}{PGP, PCKS7} \\
		Related Patterns      & \parbox[t]{0.9\columnwidth}{Message Decryptor} \\
		Known Uses            & \parbox[t]{0.9\columnwidth}{\cite{sap-hci}} \\
	\end{tabular}
	\label{tab:sec-patterns-7}
\end{table}

\begin{table}
	\caption{Encrypting Adapter \index{Encrypting Adapter|textbf} \label{ceip:encrypting-adapter}}
	\begin{tabular}{l*{2}{l}r}
		\hline
		Pattern Name          & Encrypting Adapter \index{Encrypting Adapter|textbf} \label{ceip:encrypting-adapter} \\
		\hline
		Intent                & \parbox[t]{0.9\columnwidth}{When an application sends a confidential message to another participant, its content shall only be read by the receiver.} \\
		Driving Question      & \parbox[t]{0.9\columnwidth}{How can messages be sent confidential and with message-privacy as \emph{Encrypted Message}?} \\
		Solution              & \parbox[t]{0.9\columnwidth}{Provide capabilities to encrypt the content, headers and/or attachments of a message.} \\
		Data Aspects          & \parbox[t]{0.9\columnwidth}{key store, public key, non-message generating, content modifying} \\
		Example               & \parbox[t]{0.9\columnwidth}{PGP, PCKS7} \\
		Related Patterns      & \parbox[t]{0.9\columnwidth}{Message Encryptor} \\
		Known Uses            & \parbox[t]{0.9\columnwidth}{\enquote{FileChannel} \cite{apache-flume}, \cite{sap-hci}} \\
	\end{tabular}
	\label{tab:sec-patterns-8}
\end{table}

\begin{table}
	\caption{Decrypting Adapter \index{Decrypting Adapter|textbf} \label{ceip:decrypting-adapter}}
	\begin{tabular}{l*{2}{l}r}
		\hline
		Pattern Name          & Decrypting Adapter \index{Decrypting Adapter|textbf} \label{ceip:decrypting-adapter} \\
		\hline
		Intent                & \parbox[t]{0.9\columnwidth}{When an application sends a confidential message to another participant, its content shall only be read by the receiver.} \\
		Driving Question      & \parbox[t]{0.9\columnwidth}{How can confidential messages \emph{Encrypted Message} be processed by the actual receiver?} \\
		Solution              & \parbox[t]{0.9\columnwidth}{Provide capabilities to decrypt the content, headers and/or attachments of a message.} \\
		Data Aspects          & \parbox[t]{0.9\columnwidth}{key store, private key, non-message generating, content modifying} \\
		Example               & \parbox[t]{0.9\columnwidth}{PGP, PCKS7} \\
		Related Patterns      & \parbox[t]{0.9\columnwidth}{Message Decryptor} \\
		Known Uses            & \parbox[t]{0.9\columnwidth}{\cite{sap-hci}} \\
	\end{tabular}
	\label{tab:sec-patterns-9}
\end{table}

\begin{table}
	\caption{Encrypting Store \index{Encrypting Store|textbf} \label{ceip:encrypting-store}} \label{todo}
	\begin{tabular}{l*{2}{l}r}
		\hline
		Pattern Name          & Encrypting Store \index{Encrypting Store|textbf} \label{ceip:encrypting-store} \\
		\hline
		Intent                & \parbox[t]{0.9\columnwidth}{When a message is stored confidential.} \\
		Driving Question      & \parbox[t]{0.9\columnwidth}{How can messages be stored confidential and with message-privacy as \emph{Encrypted Message}?} \\
		Solution              & \parbox[t]{0.9\columnwidth}{Message level security} \\
		Data Aspects          & \parbox[t]{0.9\columnwidth}{key store, public key, non-message generating, content modifying} \\
		Example               & \parbox[t]{0.9\columnwidth}{PGP, PCKS7} \\
		Related Patterns      & \parbox[t]{0.9\columnwidth}{Message Encryptor} \\
		Known Uses            & \parbox[t]{0.9\columnwidth}{configurable during \enquote{Persist} \cite{sap-hci}} \\
	\end{tabular}
	\label{tab:sec-patterns-10}
\end{table}

\begin{figure}[tbh]
	\centering
	\includegraphics[width=1.0\columnwidth]{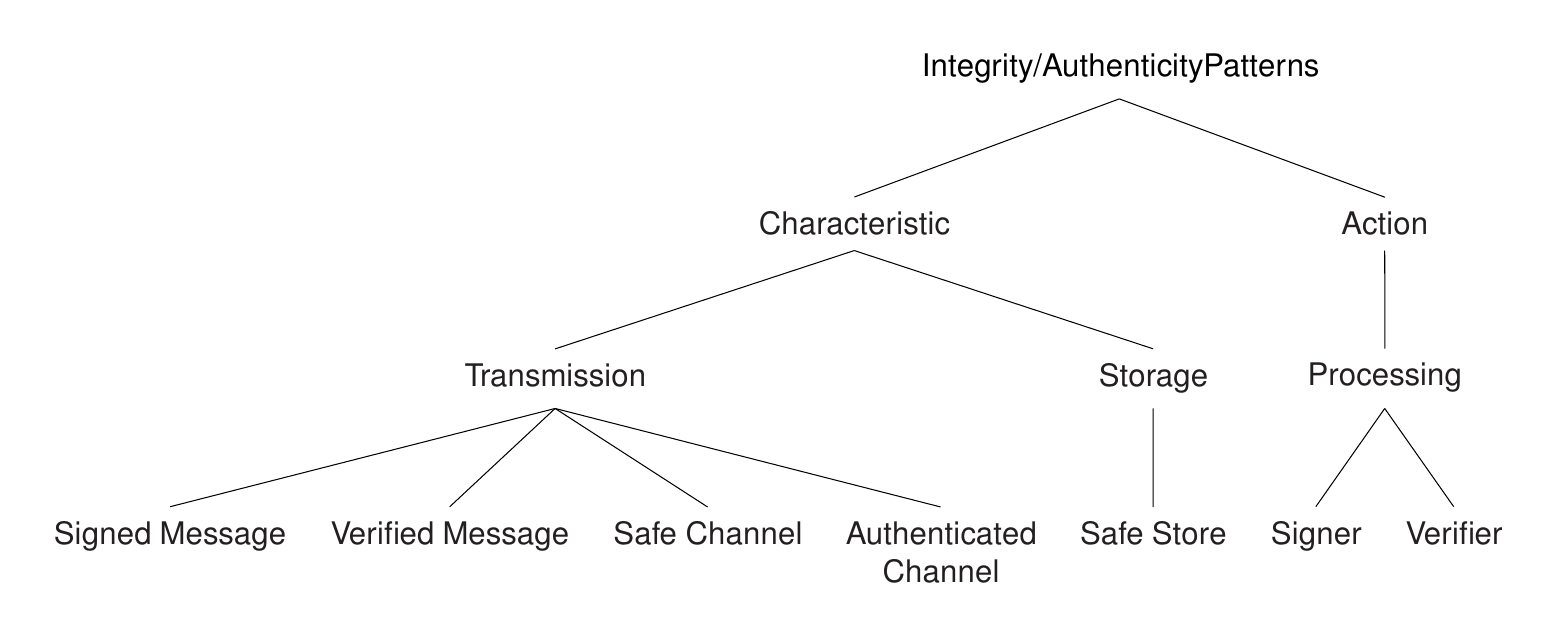}
	\caption{Integrity and Authenticity patterns grouped by characteristics and actions 
		as well as information states: processing, transmission and storage \cite{mccumber1991information}.}
	\label{fig:security_integrity_authenticity_patterns}
\end{figure}

\paragraph{Integrity and Authenticity} These patterns are collected in \cref{fig:security_integrity_authenticity_patterns} and ensure the completeness, accuracy and absence of unauthorized modifications during message processing and verify the claim of identity.
The characteristics of signed and verified messages (\cref{ceip:signed-message}, \cref{ceip:verified-message}) denote the message level and safe and authenticated channel (\cref{ceip:safe-channel}, \cref{ceip:authenticated-channel}) stand for transport level integrity and authenticity. On storage level, the pattern is called safe store (\cref{ceip:safe-store}).

The corresponding action-patterns are the signer and verifier (\cref{ceip:signer}, \cref{ceip:verifier}).



\begin{table}
	\caption{Signed Message \index{Signed Message|textbf} \label{ceip:signed-message}}
	\begin{tabular}{l*{2}{l}r}
		\hline
		Pattern Name          & Signed Message \index{Signed Message|textbf} \label{ceip:signed-message} \\
		\hline
		Intent                & \parbox[t]{0.9\columnwidth}{Ensuring a message's authenticity, integrity and non-repudiation.} \\
		Driving Question      & \parbox[t]{0.9\columnwidth}{How can a message's authenticity, integrity and non-repudiation be guaranteed?} \\
		Solution              & \parbox[t]{0.9\columnwidth}{Sign a message.} \\
		Data Aspects          & \parbox[t]{0.9\columnwidth}{certificate-based} \\
		Related Patterns      & \parbox[t]{0.9\columnwidth}{Message Signer} \\
		Known Uses            & \parbox[t]{0.9\columnwidth}{implicitly in \cite{Ibsen:2010:CA:1965487}, \cite{sap-hci}} \\
	\end{tabular}
	\label{tab:sec-patterns-16}
\end{table}

\begin{table} \label{todo}
	\caption{Verified Message \index{Verified Message|textbf} \label{ceip:verified-message}}
	\begin{tabular}{l*{2}{l}r}
		\hline
		Pattern Name          & Verified Message \index{Verified Message|textbf} \label{ceip:verified-message} \\
		\hline
		Intent                & \parbox[t]{0.9\columnwidth}{Verifying a message's authenticity, integrity and non-repudiation.} \\
		Driving Question      & \parbox[t]{0.9\columnwidth}{How can a message's authenticity, integrity and non-repudiation be verified?} \\
		Solution              & \parbox[t]{0.9\columnwidth}{Verify the signature of a message.} \\
		Data Aspects          & \parbox[t]{0.9\columnwidth}{certificate-based} \\
		Related Patterns      & \parbox[t]{0.9\columnwidth}{Signature Verifier} \\
		Known Uses            & \parbox[t]{0.9\columnwidth}{implicitly in \cite{Ibsen:2010:CA:1965487}, \cite{sap-hci}} \\
	\end{tabular}
	\label{tab:sec-patterns-17}
\end{table}

\begin{table}
	\caption{Safe Channel \index{Safe Channel|textbf} \label{ceip:safe-channel}}
	\begin{tabular}{l*{2}{l}r}
		\hline
		Pattern Name          & Safe Channel \index{Safe Channel|textbf} \\
		\hline
		Intent                & \parbox[t]{0.9\columnwidth}{Ensure integrity on transport level.} \\
		Driving Question      & \parbox[t]{0.9\columnwidth}{How to ensure integrity on transport level.} \\
		Solution              & \parbox[t]{0.9\columnwidth}{Provide integrity on transport level.} \\
		Data Aspects          & \parbox[t]{0.9\columnwidth}{certificate-based} \\
		Related Patterns      & \parbox[t]{0.9\columnwidth}{Message Signer, Authenticated Channel} \\
		Known Uses            & \parbox[t]{0.9\columnwidth}{\enquote{File channel integrity tool} \cite{apache-flume}, implicitly in \cite{sap-hci}} \\
	\end{tabular}
	\label{tab:sec-patterns-11}
\end{table}

\begin{table}
	\caption{Safe Store \index{Safe Store|textbf} \label{ceip:safe-store}}
	\begin{tabular}{l*{2}{l}r}
		\hline
		Pattern Name          & Safe Store \index{Safe Store|textbf} \\
		\hline
		Intent                & \parbox[t]{0.9\columnwidth}{Ensure integrity on storage level.} \\
		Driving Question      & \parbox[t]{0.9\columnwidth}{How to ensure integrity on storage level.} \\
		Solution              & \parbox[t]{0.9\columnwidth}{Provide integrity on transport level.} \\
		Data Aspects          & \parbox[t]{0.9\columnwidth}{certificate-based} \\
		Related Patterns      & \parbox[t]{0.9\columnwidth}{Message Signer, Safe Channel} \\
		Known Uses            & \parbox[t]{0.9\columnwidth}{implicitly in \cite{sap-hci}} \\
	\end{tabular}
	\label{tab:sec-patterns-12}
\end{table}

\begin{table}
	\caption{Authenticated Channel \index{Authenticated Channel|textbf} \label{ceip:authenticated-channel}}
	\begin{tabular}{l*{2}{l}r}
		\hline
		Pattern Name          & Authenticated Channel \index{Authenticated Channel|textbf} \\
		\hline
		Intent                & \parbox[t]{0.9\columnwidth}{Ensure authenticity on channel level.} \\
		Driving Question      & \parbox[t]{0.9\columnwidth}{How to ensure authenticity on channel level.} \\
		Solution              & \parbox[t]{0.9\columnwidth}{Ensure authenticity on transport level.} \\
		Data Aspects          & \parbox[t]{0.9\columnwidth}{certificate, user/password, token} \\
		Variations            & \parbox[t]{0.9\columnwidth}{Basic, certificate-based} \\
		Example               & \parbox[t]{0.9\columnwidth}{certificate-based, basic authentication} \\
		Related Patterns      & \parbox[t]{0.9\columnwidth}{Key Store, Trust Store, Secure Store} \\
		Known Uses            & \parbox[t]{0.9\columnwidth}{Kerberos authentication \cite{apache-flume}, two-way SSL authentication (implicit) \cite{apache-nifi}, \cite{sap-hci}} \\
	\end{tabular}
	\label{tab:sec-patterns-13}
\end{table}

\begin{table}
	\caption{Message Signer \index{Message Signer|textbf} \label{ceip:signer}}
	\begin{tabular}{l*{2}{l}r}
		\hline
		Pattern Name          & Message Signer \index{Message Signer|textbf} \label{ceip:signer} \\
		\hline
		Intent                & \parbox[t]{0.9\columnwidth}{Ensure the authenticity, integrity and non-repudation of the message content.} \\
		Driving Question      & \parbox[t]{0.9\columnwidth}{How can the authenticity of a message be ensured?} \\
		Context               & \parbox[t]{0.9\columnwidth}{Signing messages is especially important, when the communication happens via a public network (\eg cloud integration).} \\
		Solution              & \parbox[t]{0.9\columnwidth}{The pattern ensures authenticity, integrity and non-repudiation on the message-level through signing the content or parts using security mechanisms like digital signatures or envelopes. The signer cannot deny signing afterwards and any change to the signed message parts can be detected using, \eg the \emph{Verifier} pattern. \emph{Message Translator} patterns might invalidate the signing.} \\
		Data Aspects          & \parbox[t]{0.9\columnwidth}{in: any message, out: message with signed content or parts; requires private key, key store} \\
		Result                & \parbox[t]{0.9\columnwidth}{The signing is done by asymmetric cryptography that requires for example the creation of a signature using a signing algorithm with the private key stored in a secured key store.} \\
		Example               & \parbox[t]{0.9\columnwidth}{PGP, PCKS7} \\
		Related Patterns      & \parbox[t]{0.9\columnwidth}{Key Store, Trust Store} \\
		Known Uses            & \parbox[t]{0.9\columnwidth}{\enquote{Camel Crypto} \cite{Ibsen:2010:CA:1965487}, \enquote{Message Signer} \cite{sap-hci}} \\
	\end{tabular}
	\label{tab:sec-patterns-14}
\end{table}


\begin{table}
	\caption{Signature Verifier \index{Signature Verifier|textbf} \label{ceip:verifier}}
	\begin{tabular}{l*{2}{l}r}
		\hline
		Pattern Name          & Signature Verifier \index{Signature Verifier|textbf} \\
		\hline
		Intent                & \parbox[t]{0.9\columnwidth}{Verify a message's authenticity, integrity and non-repudiation} \\
		Driving Question      & \parbox[t]{0.9\columnwidth}{How to verify a message's authenticity, integrity and non-repudiation?} \\
		Solution              & \parbox[t]{0.9\columnwidth}{Provide certificate-based verification on message level.} \\
		Data Aspects          & \parbox[t]{0.9\columnwidth}{in: Signed Message, out: Signed Message (read-only); exception: verification failed; requires access: certificate, Key Store, non-message generating, read-only} \\
		Variations            & \parbox[t]{0.9\columnwidth}{Signature Verifier, XMLSignatureVerifier} \\
		Related Patterns      & \parbox[t]{0.9\columnwidth}{Message Signer, Signed Message, Key Store, Trust Store} \\
		Known Uses            & \parbox[t]{0.9\columnwidth}{\enquote{Verifier} \cite{sap-hci}} \\
	\end{tabular}
	\label{tab:sec-patterns-15}
\end{table}


\paragraph{Availability} The availability of resources like \emph{Message Store} \cite{Hohpe:2003:EIP:940308}, \emph{Key Store} or \emph{Data Store} are crucial. \Cref{ceip:redundant-store} defines the countermeasure as a redundant store pattern.

\begin{table}
	\caption{Redundant Store \index{Redundant Store|textbf} \label{ceip:redundant-store}}
	\begin{tabular}{l*{2}{l}r}
		\hline
		Pattern Name          & Redundant Store \index{Redundant Store|textbf} \\
		\hline
		Intent                & \parbox[t]{0.9\columnwidth}{Ensure availability of storage resources.} \\
		Driving Question      & \parbox[t]{0.9\columnwidth}{How to ensure availability of storage resources?} \\
		Solution              & \parbox[t]{0.9\columnwidth}{Provide redundant hardware, service (probably even high-available and disaster recoverable)} \\
		Data Aspects          & \parbox[t]{0.9\columnwidth}{availability level: HA, DR} \\
		Variations            & \parbox[t]{0.9\columnwidth}{Failover, HA, DR} \\
		Related Patterns      & \parbox[t]{0.9\columnwidth}{Message Store, Data Store, Key Store, Trust Store, Secure Store} \\
		Known Uses            & \parbox[t]{0.9\columnwidth}{\enquote{MorphlineSolrSinks}, \enquote{Kafka Channel} \cite{apache-flume}, implicitly in \cite{sap-hci}; \emph{Tandem Store} \cite{Hafiz:2012:GPL:2384592.2384607,blakley2004}} \\
	\end{tabular}
	\label{tab:sec-patterns-18}
\end{table}


\paragraph{Non-repudiation / Auditability / Accountability}
Being able to prove which event happenend during message processing when and with which privileges (role) and by whom (user) is important for detecting security issues as well as using the information for metering.


\begin{table}
	\caption{Audit Log \index{Audit Log|textbf} \label{ceip:audit-log}}
	\begin{tabular}{l*{2}{l}r}
		\hline
		Pattern Name          & Audit Log \index{Audit Log|textbf} \\
		\hline
		Intent                & \parbox[t]{0.9\columnwidth}{Keep a record of those events you consider relevant for (security) auditing: accountability, non-repudiation, auditability} \\
		Driving Question      & \parbox[t]{0.9\columnwidth}{How to keep a record of those events considered relevant for (security) auditing.} \\
		Solution              & \parbox[t]{0.9\columnwidth}{Provide a secure logging capability, which tracks events and (especially) all its own configurations.} \\
		Data Aspects          & \parbox[t]{0.9\columnwidth}{Log entries} \\
		Example               & \parbox[t]{0.9\columnwidth}{Record event-based endpoints, unsuccessful message send events, changes to the audit log.} \\
		Related Patterns      & \parbox[t]{0.9\columnwidth}{Monitor} \\
		Known Uses            & \parbox[t]{0.9\columnwidth}{implicitly in \cite{sap-hci}} \\
	\end{tabular}
	\label{tab:sec-patterns-19}
\end{table}

\paragraph{Authorization, Non-Delegation}


\begin{table}
	\caption{Token \index{Token|textbf} \label{ceip:token}}
	\begin{tabular}{l*{2}{l}r}
		\hline
		Pattern Name          & Token \index{Token|textbf} \label{ceip:token} \\
		\hline
		Intent                & \parbox[t]{0.9\columnwidth}{Grant role-based access to a user.} \\
		Driving Question      & \parbox[t]{0.9\columnwidth}{How to grant role-based access to a user?} \\
		Solution              & \parbox[t]{0.9\columnwidth}{Provide secure token, which is passed as part of each during conversation.} \\
		Example               & \parbox[t]{0.9\columnwidth}{OAuth} \\
		Related Patterns      & \parbox[t]{0.9\columnwidth}{Secure Store} \\
		Known Uses            & \parbox[t]{0.9\columnwidth}{\enquote{GetTwitter} \cite{apache-nifi}, \enquote{Secure Paramters} in SAP HCI ADK \cite{sap-hci}} \\
	\end{tabular}
	\label{tab:sec-patterns-20}
\end{table}


\begin{table}
	\caption{Expiring Token \index{Expiring Token|textbf} \label{ceip:expiring-token}}
	\begin{tabular}{l*{2}{l}r}
		\hline
		Pattern Name          & Expiring Token \index{Expiring Token|textbf} \\
		\hline
		Intent                & \parbox[t]{0.9\columnwidth}{Improve security for role-based access of users via Tokens.} \\
		Driving Question      & \parbox[t]{0.9\columnwidth}{How to improve security, when granting role-based access to a user via Tokens?} \\
		Solution              & \parbox[t]{0.9\columnwidth}{Reduce the validity of a Token through expiration.} \\
		Data Aspects          & \parbox[t]{0.9\columnwidth}{Token} \\
		Example               & \parbox[t]{0.9\columnwidth}{OAuth} \\
		Related Patterns      & \parbox[t]{0.9\columnwidth}{Secure Store} \\
		Known Uses            & \parbox[t]{0.9\columnwidth}{\enquote{GetTwitter} \cite{apache-nifi}, \enquote{Secure Paramters} in SAP HCI ADK \cite{sap-hci}} \\
	\end{tabular}
	\label{tab:sec-patterns-21}
\end{table}

\begin{table}
	\caption{Refresh Token \index{Refresh Token|textbf} \label{ceip:refresh-token}}
	\begin{tabular}{l*{2}{l}r}
		\hline
		Pattern Name          & Refresh Token \index{Refresh Token|textbf} \\
		\hline
		Intent                & \parbox[t]{0.9\columnwidth}{Deal with expiring tokens.} \\
		Driving Question      & \parbox[t]{0.9\columnwidth}{How to deal with expiring tokens?} \\
		Solution              & \parbox[t]{0.9\columnwidth}{Allow to (automatically) re-negotiate expiring tokens?} \\
		Data Aspects          & \parbox[t]{0.9\columnwidth}{Expiring Token} \\
		Example               & \parbox[t]{0.9\columnwidth}{OAuth} \\
		Related Patterns      & \parbox[t]{0.9\columnwidth}{Secure Store} \\
		Known Uses            & \parbox[t]{0.9\columnwidth}{\enquote{GetTwitter} \cite{apache-nifi}, \enquote{Secure Paramters} in SAP HCI ADK \cite{sap-hci}} \\
	\end{tabular}
	\label{tab:sec-patterns-22}
\end{table}

\begin{table}
	\caption{Token-based Authorizer \index{Token-based Authorizer|textbf} \label{ceip:token-authorizer}}
	\begin{tabular}{l*{2}{l}r}
		\hline
		Pattern Name          & Token-based Authorizer \index{Token-based Authorizer|textbf} \\
		\hline
		Intent                & \parbox[t]{0.9\columnwidth}{Validate token-based authorizations.} \\
		Driving Question      & \parbox[t]{0.9\columnwidth}{How to validate token-based authorizations?} \\
		Solution              & \parbox[t]{0.9\columnwidth}{Provide authorization checks for messages with tokens.} \\
		Data Aspects          & \parbox[t]{0.9\columnwidth}{Secure Tokens} \\
		Example               & \parbox[t]{0.9\columnwidth}{OAuth} \\
		Related Patterns      & \parbox[t]{0.9\columnwidth}{Secure Store, Adapter, Message Endpoint} \\
		Known Uses            & \parbox[t]{0.9\columnwidth}{Pluggable Authorization \cite{apache-nifi}, SAP Social Media ADK adapters \cite{sap-hci}} \\
	\end{tabular}
	\label{tab:sec-patterns-22}
\end{table}



\begin{table}
	\caption{Principle Propagation \index{Principle Propagation|textbf} \label{ceip:propagation}}
	\begin{tabular}{l*{2}{l}r}
		\hline
		Pattern Name          & Principle Propagation \index{Principle Propagation|textbf} \\
		\hline
		Intent                & \parbox[t]{0.9\columnwidth}{} \\
		Driving Question      & \parbox[t]{0.9\columnwidth}{How to delegate authorizations to a third-party?} \\
		Solution              & \parbox[t]{0.9\columnwidth}{Allow authorization forwarding.} \\
		Data Aspects          & \parbox[t]{0.9\columnwidth}{Token} \\
		Related Patterns      & \parbox[t]{0.9\columnwidth}{Token} \\
		Known Uses            & \parbox[t]{0.9\columnwidth}{\enquote{Principle Propagation} \cite{sap-pi}} \\
	\end{tabular}
	\label{tab:sec-patterns-23}
\end{table}


\newpage
\subsection{Exception Patterns}
\label{sec:exception-patterns}

\begin{figure}[tbh]
	\centering
	\includegraphics[width=0.9\columnwidth]{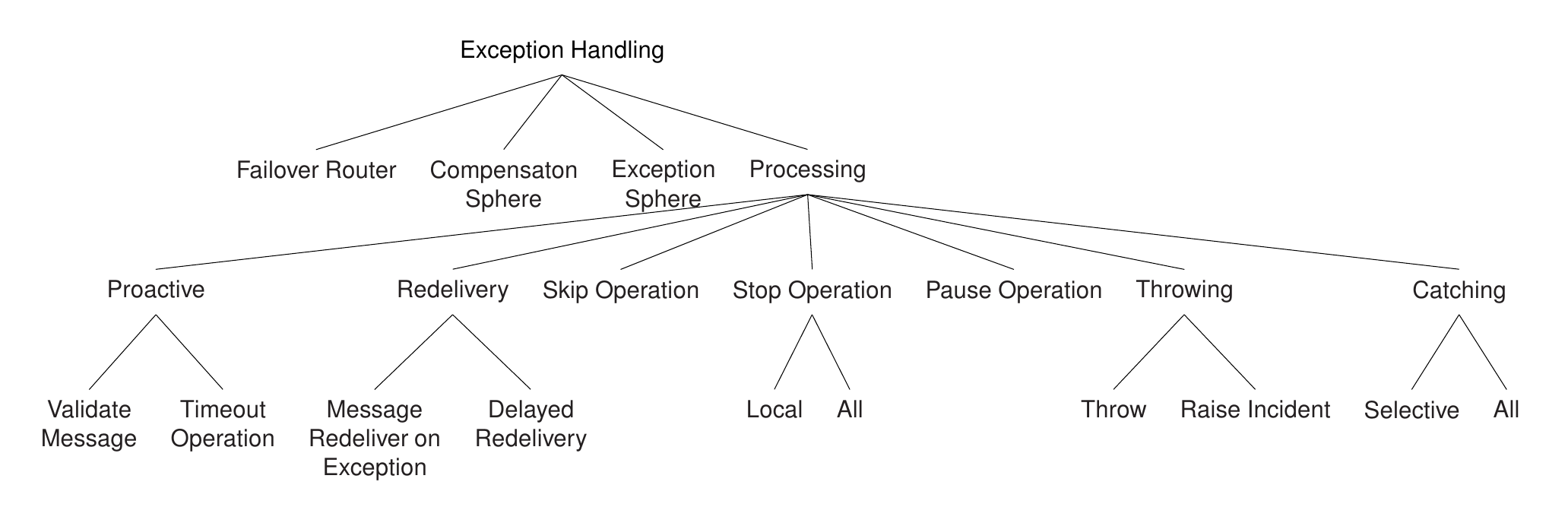}
	\caption{Exception Handling Patterns.}
	\label{fig:exception-overview}
\end{figure}

The exception patterns were introduced and analyzed in \cite{DBLP:conf/edoc/RitterS14}, thus summarized in \cref{fig:exception-overview} and not further discussed:

\begin{itemize}
	\item Failover Router\index{Failover Router|textbf}
	\item Compensation Sphere\index{Compensation Sphere|textbf}
	\item Exception Sphere\index{Exception Sphere|textbf}
	\item Validate Message\index{Validate Message|textbf}
	\item Timeout Operation\index{Timeout Operation|textbf}
	\item Message Redelivery on Exception\index{Message Redelivery on Exception|textbf}
	\item Delayed Redelivery\index{Delayed Redelivery|textbf}
	\item Skip Operation\index{Skip Operation|textbf}
	\item Stop Operation Local\index{Stop Operation Local|textbf}
	\item Stop Operation All\index{Stop Operation All|textbf}
	\item Pause Operation\index{Pause Operation|textbf}
	\item Throw\index{Throw|textbf}
	\item Raise Incident\index{Raise Incident|textbf} (similar to monitoring pattern \cref{sec:monitoring-pattern})
	\item Catch Selective\index{Catch Selective|textbf}
	\item Catch All\index{Catch All|textbf}
\end{itemize}

\newpage
\subsection{Monitoring and Operation Patterns}
\label{sec:monitoring-pattern}

The current EIP cover basic monitoring and operation patterns \cite{Hohpe:2003:EIP:940308}: \emph{Control Bus} for administrating the messaging system, \emph{Wire Tab} for routing message copy \eg to a \emph{Message Store} for monitoring, \emph{Message History} for provenance, and \emph{Smart Proxies} for asynchronous message tracking.

\begin{figure}[tbh]
	\centering
	\includegraphics[width=0.9\columnwidth]{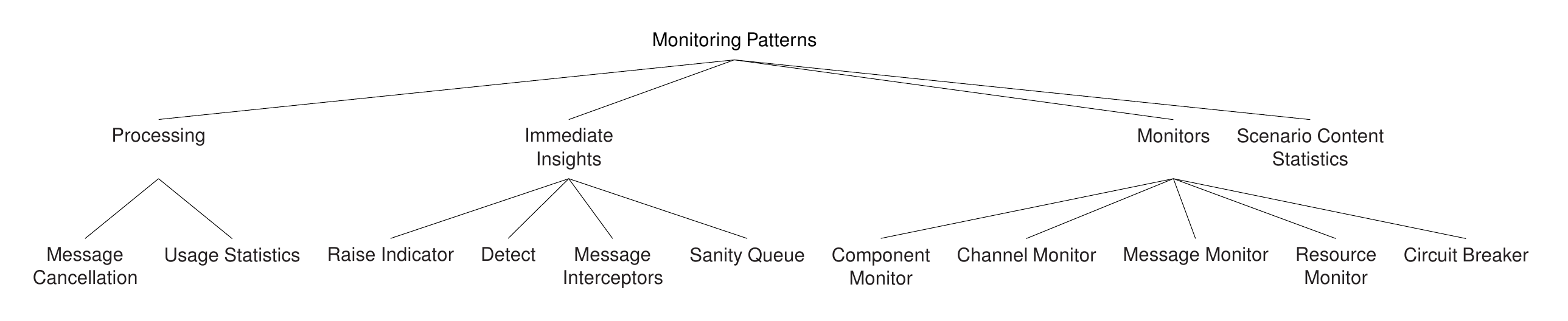}
	\caption{Monitoring Patterns.}
	\label{fig:monitoring-overview}
\end{figure}

In addition, there are more aspects to monitoring and operations that can be found in current integration system implementations. Some of them are collected subsequently (others like the circuit breaker \cref{sec:adapter-pattern} can be found in related sections). \Cref{fig:monitoring-overview} gives an overview of the collected monitoring patterns.

The processing of messages might require their cancellation (cf. \cref{tab:monitoring-patterns-1}), \eg based on defined usage statistics (cf. \cref{tab:monitoring-patterns-3})).

\begin{table}
	\caption{Message Cancellation \index{Message Cancellation|textbf} \label{ceip:message_cancel}}
	\begin{tabular}{l*{2}{l}r}
		\hline
		Pattern Name          & Message Cancellation\index{Message Cancellation|textbf} \\
		\hline
		Intent                & \parbox[t]{0.9\columnwidth}{Cancel the processing of a message.} \\
		Driving Question      & \parbox[t]{0.9\columnwidth}{How to cancel the processing of a message?} \\
		Solution              & \parbox[t]{0.9\columnwidth}{Define a condition under which the message will not be processed further.} \\
		Data Aspects          & \parbox[t]{0.9\columnwidth}{condition} \\
		Related Patterns      & \parbox[t]{0.9\columnwidth}{Message Expiration \cite{Hohpe:2003:EIP:940308}, Stop Operation (Local,All) \cite{DBLP:conf/edoc/RitterS14}, Validate Message \cite{DBLP:conf/edoc/RitterS14}} \\
		Known Uses            & \parbox[t]{0.9\columnwidth}{\cite{sap-hci}} \\
	\end{tabular}
	\label{tab:monitoring-patterns-1}
\end{table}



\begin{table}
	\caption{Usage Statistics \index{Usage Statistics|textbf} \label{ceip:usage}}
	\begin{tabular}{l*{2}{l}r}
		\hline
		Pattern Name          & Usage Statistics\index{Usage Statistics|textbf} \\
		\hline
		Intent                & \parbox[t]{0.9\columnwidth}{A measure for the usage of components.} \\
		Driving Question      & \parbox[t]{0.9\columnwidth}{How to measure the usage of components during message processing?} \\
		Solution              & \parbox[t]{0.9\columnwidth}{Define performance indicators for components.} \\
		Data Aspects          & \parbox[t]{0.9\columnwidth}{usage indicators, persistent} \\
		Related Patterns      & \parbox[t]{0.9\columnwidth}{Scenario Content Statistics} \\
		Known Uses            & \parbox[t]{0.9\columnwidth}{\cite{sap-hci}} \\
	\end{tabular}
	\label{tab:monitoring-patterns-3}
\end{table}

While the whole message processing and failure handling shall be automatic, there might be situations (\eg due to business semantics), which require raising an indication (cf. \cref{tab:monitoring-patterns-4}) for manual post-processing or information. Sometimes this can be preceded by the detection of a special situation (cf. \cref{tab:monitoring-patterns-5}). The information can be collected by interceptors (cf. \cref{tab:monitoring-patterns-6}) and provided asynchronously by a sanity queue (cf. \cref{tab:monitoring-patterns-7}).

\begin{table}
	\caption{Raise Indicator \index{Raise Indicator|textbf} \label{ceip:alert}}
	\begin{tabular}{l*{2}{l}r}
		\hline
		Pattern Name          & Raise Indicator\index{Raise Incident|textbf} \\
		\hline
		Intent                & \parbox[t]{0.9\columnwidth}{Inform about an important event during message processing.} \\
		Driving Question      & \parbox[t]{0.9\columnwidth}{How to inform about an important event during message processing?} \\
		Solution              & \parbox[t]{0.9\columnwidth}{Raise an indicator.} \\
		Data Aspects          & \parbox[t]{0.9\columnwidth}{persistent} \\
		Example               & \parbox[t]{0.9\columnwidth}{Send e-mail, Monitor} \\
		Related Patterns      & \parbox[t]{0.9\columnwidth}{Detect} \\
		Known Uses            & \parbox[t]{0.9\columnwidth}{\enquote{Indicator} \cite{apache-nifi}, \enquote{Alert} \cite{sap-hci}} \\
	\end{tabular}
	\label{tab:monitoring-patterns-4}
\end{table}

\begin{table}
	\caption{Detect \index{Detect|textbf} \label{ceip:monitor-detect}}
	\begin{tabular}{l*{2}{l}r}
		\hline
		Pattern Name          & Detect \index{Detect|textbf} \\
		\hline
		Intent                & \parbox[t]{0.9\columnwidth}{Detect inactive, active components like channel or pattern/activity.} \\
		Driving Question      & \parbox[t]{0.9\columnwidth}{How to detect inactive components like channels or patterns?} \\
		Solution              & \parbox[t]{0.9\columnwidth}{Send out an alert when a component did not have any data for a specified amount of time.} \\
		Data Aspects          & \parbox[t]{0.9\columnwidth}{conditions} \\
		Related Patterns      & \parbox[t]{0.9\columnwidth}{Usage statistics, Channel Monitor, Component Monitor} \\
		Known Uses            & \parbox[t]{0.9\columnwidth}{\enquote{Monitor Activity} \cite{apache-nifi}} \\
	\end{tabular}
	\label{tab:monitoring-patterns-5}
\end{table}

\begin{table}
	\caption{Message Interceptor \index{Message Interceptor|textbf} \label{ceip:interceptor}}
	\begin{tabular}{l*{2}{l}r}
		\hline
		Pattern Name          & Message Interceptor \index{Message Interceptor|textbf} \\
		\hline
		Intent                & \parbox[t]{0.9\columnwidth}{Gather usage statistics on channel level.} \\
		Driving Question      & \parbox[t]{0.9\columnwidth}{How to gather usage statistics on channel level?} \\
		Solution              & \parbox[t]{0.9\columnwidth}{Inject a configurable \enquote{listener} between components.} \\
		Data Aspects          & \parbox[t]{0.9\columnwidth}{conditions, read-only, non-message creating} \\
		Related Patterns      & \parbox[t]{0.9\columnwidth}{Usage statistics, Wire Tap \cite{Hohpe:2003:EIP:940308}, Intercepting Validator \cite{Hafiz:2012:GPL:2384592.2384607,steel2005core}} \\
		Known Uses            & \parbox[t]{0.9\columnwidth}{\enquote{Interceptor} \cite{Ibsen:2010:CA:1965487}, \enquote{Interceptor} \cite{apache-flume}} \\
	\end{tabular}
	\label{tab:monitoring-patterns-6}
\end{table}


\begin{table}
	\caption{Sanity Queue \index{Sanity Queue|textbf} \label{ceip:sanityqueue}}
	\begin{tabular}{l*{2}{l}r}
		\hline
		Pattern Name          & Sanity Queue \index{Sanity Queue|textbf} \\
		\hline
		Intent                & \parbox[t]{0.9\columnwidth}{Stay informed about a system's sanity.} \\
		Driving Question      & \parbox[t]{0.9\columnwidth}{How to stay informed about the system's sanity?} \\
		Solution              & \parbox[t]{0.9\columnwidth}{Create sanity queues for important aspects and register on events.} \\
		Data Aspects          & \parbox[t]{0.9\columnwidth}{queues, persistent} \\
		Related Patterns      & \parbox[t]{0.9\columnwidth}{Wire Tap} \\
		Known Uses            & \parbox[t]{0.9\columnwidth}{JMS sanity check} \\
	\end{tabular}
	\label{tab:monitoring-patterns-7}
\end{table}

The respective events and incidents can be tracked by monitors, \eg for components (cf. \cref{tab:monitoring-patterns-8}), channels (cf. \cref{tab:monitoring-patterns-9}) and messages (cf. \cref{tab:monitoring-patterns-10}). In addition, the system resource information (cf. \cref{tab:monitoring-patterns-11}) as well as the overview of all open circuits (cf. \cref{tab:endpoint-patterns-1}) might be of interest.
The monitoring of hybrid applications and integration sceanrios is covered by \cref{tab:monitoring-patterns-112}.

\begin{table}
	\caption{Component Monitor \index{Component Monitor|textbf} \label{ceip:component_monitor}}
	\begin{tabular}{l*{2}{l}r}
		\hline
		Pattern Name          & Component Monitor \index{Component Monitor|textbf} \\
		\hline
		Intent                & \parbox[t]{0.9\columnwidth}{Measure usage statistics and behavior of a component.} \\
		Driving Question      & \parbox[t]{0.9\columnwidth}{How to measure usage statistics and behavior of a component?} \\
		Solution              & \parbox[t]{0.9\columnwidth}{Monitor configurable characteristics of a component and store them persistently as statistics.} \\
		Data Aspects          & \parbox[t]{0.9\columnwidth}{Configurations, statistical records, KPIs, persistent} \\
		Example               & \parbox[t]{0.9\columnwidth}{Monitor throughput, exceptions and raise indicator or load balance, if KPIs are not fulfilled.} \\
		Related Patterns      & \parbox[t]{0.9\columnwidth}{Stop Operation, Pause Operation (both from \cite{DBLP:conf/edoc/RitterS14}), Failover Router \cite{DBLP:conf/edoc/RitterS14}, Data Store, Smart Proxy \cite{Hohpe:2003:EIP:940308}} \\
		Known Uses            & \parbox[t]{0.9\columnwidth}{\enquote{Adapter Monitor} \cite{hystrix}, conceptually supprted by \cite{Nygard07}} \\
	\end{tabular}
	\label{tab:monitoring-patterns-8}
\end{table}

\begin{table}
	\caption{Channel Monitor \index{Channel Monitor|textbf} \label{ceip:adapter_monitor}}
	\begin{tabular}{l*{2}{l}r}
		\hline
		Pattern Name          & Channel Monitor \index{Channel Monitor|textbf} \\
		\hline
		Intent                & \parbox[t]{0.9\columnwidth}{Measure communication with endpoints.} \\
		Driving Question      & \parbox[t]{0.9\columnwidth}{How to measure success, failures (exceptions thrown by endpoint), timeouts for requests?} \\
		Solution              & \parbox[t]{0.9\columnwidth}{Monitor configurable characteristics of the communication with endpoints and store them persistently as statistics.} \\
		Data Aspects          & \parbox[t]{0.9\columnwidth}{Configurations, statistical records, KPIs, persistent} \\
		Example               & \parbox[t]{0.9\columnwidth}{Monitor timeout to endpoints and decide to apply a Circuit Breaker pattern.} \\
		Related Patterns      & \parbox[t]{0.9\columnwidth}{Circuit Breaker, Timeout Synchronous Request, Failover Router \cite{DBLP:conf/edoc/RitterS14}, Command, Data Store, Smart Proxy \cite{Hohpe:2003:EIP:940308}} \\
		Known Uses            & \parbox[t]{0.9\columnwidth}{\enquote{Adapter Monitor} \cite{hystrix}, conceptually supported by \cite{Nygard07}} \\
	\end{tabular}
	\label{tab:monitoring-patterns-9}
\end{table}

\begin{table}
	\caption{Message Monitor \index{Message Monitor|textbf} \label{ceip:message_monitor}}
	\begin{tabular}{l*{2}{l}r}
		\hline
		Pattern Name          & Message Monitor \index{Message Monitor|textbf} \\
		\hline
		Intent                & \parbox[t]{0.9\columnwidth}{Monitor the states of all processed messages.} \\
		Driving Question      & \parbox[t]{0.9\columnwidth}{How to monitor the states of all currently processed messages?} \\
		Solution              & \parbox[t]{0.9\columnwidth}{Provide a status monitor that shows the Message History.} \\
		Data Aspects          & \parbox[t]{0.9\columnwidth}{message provenance data} \\
		Related Patterns      & \parbox[t]{0.9\columnwidth}{Message History} \\
		Known Uses            & \parbox[t]{0.9\columnwidth}{\enquote{Message Processing Log}\cite{sap-hci}} \\
	\end{tabular}
	\label{tab:monitoring-patterns-10}
\end{table}

\begin{table}
	\caption{Resource Monitor \index{Resource Monitor|textbf} \label{ceip:resource_monitor}}
	\begin{tabular}{l*{2}{l}r}
		\hline
		Pattern Name          & Resource Monitor \index{Resource Monitor|textbf} \\
		\hline
		Intent                & \parbox[t]{0.9\columnwidth}{Monitor the system resources.} \\
		Driving Question      & \parbox[t]{0.9\columnwidth}{How to monitor the system resources and react on critical situations?} \\
		Solution              & \parbox[t]{0.9\columnwidth}{Provide a status monitor that shows the system's recources and allow the definition of thresholds for raising indicators.} \\
		Data Aspects          & \parbox[t]{0.9\columnwidth}{resource statistics} \\
		Example               & \parbox[t]{0.9\columnwidth}{Memory consumption, CPU and network load} \\
		Related Patterns      & \parbox[t]{0.9\columnwidth}{Raise indicator} \\
		Known Uses            & \parbox[t]{0.9\columnwidth}{\enquote{MonitorDiskUsage}, \enquote{MonitorMemory} \cite{apache-nifi}} \\
	\end{tabular}
	\label{tab:monitoring-patterns-11}
\end{table}

\begin{table}
	\caption{Hybrid Monitor \index{Hybrid Monitor|textbf} \label{ceip:hybrid_monitor}}
	\begin{tabular}{l*{2}{l}r}
		\hline
		Pattern Name          & Hybrid Monitor \index{Hybrid Monitor|textbf} \\
		\hline
		Intent                & \parbox[t]{0.9\columnwidth}{Monitor across different integration platforms.} \\
		Driving Question      & \parbox[t]{0.9\columnwidth}{How to monitor integration scenarios across different integration platforms?} \\
		Solution              & \parbox[t]{0.9\columnwidth}{Provide a monitor that combines (data ingestion) different, vendor-specific monitoring data, normalizes it (\ie format, timings), calculates and visualizes combined metrics.} \\
		Data Aspects          & \parbox[t]{0.9\columnwidth}{heterogeneous monitoring data} \\
		Example               & \parbox[t]{0.9\columnwidth}{Cross-Cloud, On-Demand to Cloud} \\
		Related Patterns      & \parbox[t]{0.9\columnwidth}{-} \\
		Known Uses            & \parbox[t]{0.9\columnwidth}{\enquote{sap-hci}} \\
	\end{tabular}
	\label{tab:monitoring-patterns-112}
\end{table}

For analytical reasons, it might be important to have statistics on the used components and adapters (cf. \cref{tab:monitoring-patterns-12}).

\begin{table}
	\caption{Scenario Content Statistics \index{Scenario Content Statistics|textbf} \label{ceip:content_analysis}}
	\begin{tabular}{l*{2}{l}r}
		\hline
		Pattern Name          & Scenario Content Statistics \index{Scenario Content Statistics|textbf} \\
		\hline
		Intent                & \parbox[t]{0.9\columnwidth}{A measure for the usage of content.} \\
		Driving Question      & \parbox[t]{0.9\columnwidth}{How to measure the usage of content within integration scenarios?} \\
		Solution              & \parbox[t]{0.9\columnwidth}{Define performance indicators for the usage of content.} \\
		Data Aspects          & \parbox[t]{0.9\columnwidth}{content statistics} \\
		Example               & \parbox[t]{0.9\columnwidth}{Number of used adapters and their configuration} \\
		Related Patterns      & \parbox[t]{0.9\columnwidth}{Usage statistics} \\
		Known Uses            & \parbox[t]{0.9\columnwidth}{\cite{sap-hci}} \\
	\end{tabular}
	\label{tab:monitoring-patterns-12}
\end{table}

\begin{figure}[tbh]
	\centering
	\includegraphics[width=0.9\columnwidth]{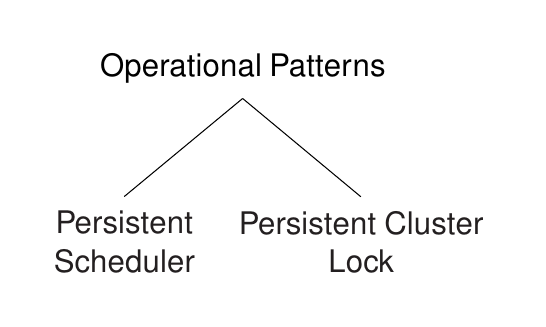}
	\caption{Operational Patterns.}
	\label{fig:operations-overview}
\end{figure}

The operational aspects mostly come from reliability and distributed system requirements as shown in \cref{fig:operations-overview}.

Upon them are the persistent scheduling (cf. \cref{tab:ops-patterns-1}) and the persistent cluster lock (cf. \cref{tab:ops-patterns-1}).

\begin{table}
	\caption{(Persistent) Scheduler \index{Scheduler|textbf} \label{ceip:scheduler}}
	\begin{tabular}{l*{2}{l}r}
		\hline
		Pattern Name          & (Persistent) Scheduler \index{Scheduler|textbf} \\
		\hline
		Intent                & \parbox[t]{0.9\columnwidth}{Start from last message or action after restart.} \\
		Driving Question      & \parbox[t]{0.9\columnwidth}{How to start scheduling from the last message or action after system restart?} \\
		Solution              & \parbox[t]{0.9\columnwidth}{Persistently store the state of the scheduler.} \\
		Data Aspects          & \parbox[t]{0.9\columnwidth}{persistent} \\
		Related Patterns      & \parbox[t]{0.9\columnwidth}{Data Store, Polling Consumer} \\
		Known Uses            & \parbox[t]{0.9\columnwidth}{\cite{Ibsen:2010:CA:1965487}, \cite{sap-hci}, conceptually supported in \cite{DBLP:phd/de/Bohm2010}} \\
	\end{tabular}
	\label{tab:ops-patterns-1}
\end{table}

\begin{table}
	\caption{(Persistent) Cluster Lock \index{Cluster Lock|textbf} \label{ceip:cluster_lock}}
	\begin{tabular}{l*{2}{l}r}
		\hline
		Pattern Name          & (Persistent) Cluster Lock \index{Cluster Lock|textbf} \\
		\hline
		Intent                & \parbox[t]{0.9\columnwidth}{Prevent several integration scenarios from processing the same information concurrently.} \\
		Driving Question      & \parbox[t]{0.9\columnwidth}{How to prevent several integration scenarios from processing the same information concurrently?} \\
		Solution              & \parbox[t]{0.9\columnwidth}{Set persistent lock.} \\
		Data Aspects          & \parbox[t]{0.9\columnwidth}{persitent} \\
		Related Patterns      & \parbox[t]{0.9\columnwidth}{Data Store, Competing Consumers \cite{Hohpe:2003:EIP:940308} } \\
		Known Uses            & \parbox[t]{0.9\columnwidth}{\cite{sap-hci}} \\
	\end{tabular}
	\label{tab:ops-patterns-2}
\end{table}

\newpage
\subsection{Endpoint and Adapter Patterns}
\label{sec:adapter-pattern}

Many of the channel access and endpoint patterns for connecting applications with messaging systems (\ie \emph{Message Channel}, \emph{Channel Adapter}, \emph{Message Endpoint}, \emph{Durable Subscriber}, \emph{Channel Purger}, \emph{Transactional Client}, \emph{Selective Consumer}, \emph{Polling Consumer}, \emph{Message Dispatcher}, \emph{Event-Driven Consumer}), and for involving existing systems in the message exchange (\ie \emph{Envelope Wrapper}, \emph{Messaging Bridge}) are covered by \cite{Hohpe:2003:EIP:940308}.

\begin{figure}[tbh]
	\centering
	\includegraphics[width=0.9\columnwidth]{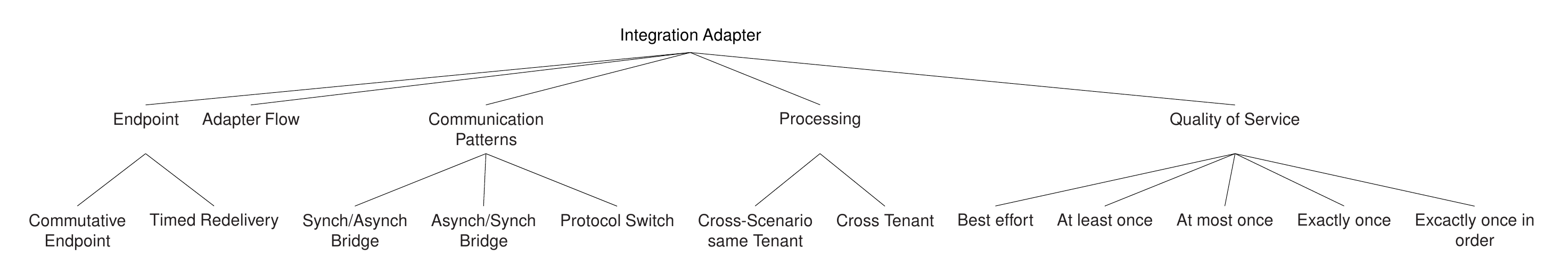}
	\caption{Integration Adapter and QoS Patterns.}
	\label{fig:adapter-overview}
\end{figure}

In addition, \cite{ritter2015integration} collected some more integration adapter and quality of service (QoS) related patterns, shown in \cref{fig:adapter-overview}, which are not further discussed here:

\begin{itemize}
	\item Commutative Endpoint\index{Commutative Endpoint|textbf}
	\item Timed Redelivery\index{Timed Redelivery|textbf}
	\item Adapter Flow\index{Adapter Flow|textbf}
	\item Synch/Asynch Bridge\index{Synch/Asynch Bridge|textbf}
	\item Asynch/Synch Bridge\index{Asynch/Synch Bridge|textbf}
	\item Protocol Switch\index{Protocol Switch|textbf}
	\item Cross scenario Processing\index{Cross scenario Processing|textbf}
	\item Cross tenant Processing\index{Cross tenant Processing|textbf}
	\item Best Effort Processing\index{Best Effort Processing|textbf}
	\item At least once Processing\index{At least once Processing|textbf}
	\item At most once Processing\index{At most once Processing|textbf}
	\item Exactly once Processing\index{Exactly once Processing|textbf}
	\item Exactly once in order Processing\index{Exactly once in order Processing|textbf}
\end{itemize}

The two patterns not discussed in \cite{ritter2015integration} are the \emph{Commutative Receiver}, and the \emph{Timed Redelivery}.
These patterns are considered endpoint or application patterns and move integration logic to the applications.
Hence they complement the existing the \emph{Idempotent Endpoint} pattern \cite{Hohpe:2003:EIP:940308} to ensure in-order and at-least-once message delivery semantics outside of the integration system.

\begin{table}
	\caption{Commutative Receiver \index{Commutative Receiver|textbf} \label{ceip:commutative_receiver}}
	\begin{tabular}{l*{2}{l}r}
		\hline
		Pattern Name          & Commutative Receiver \index{Commutative Receiver|textbf} \\
		\hline
		Intent                & \parbox[t]{0.9\columnwidth}{Enable the endpoint/application to deal with out-of-order messages.} \\
		Driving Question      & \parbox[t]{0.9\columnwidth}{How to ensure in-order processing without intermediate storage, \eg in form of a resequencer \cite{Hohpe:2003:EIP:940308}?} \\
		Context               & \parbox[t]{0.9\columnwidth}{Out-of-order communication with endpoints/applications} \\
		Solution              & \parbox[t]{0.9\columnwidth}{Guarantee that endpoint/application handle arriving out-of-order messages will be stored and then processed in the correct order.} \\
		Result                & \parbox[t]{0.9\columnwidth}{Handles out of order messages and applies them in-order.} \\
		Related Patterns      & \parbox[t]{0.9\columnwidth}{Resequencer \cite{Hohpe:2003:EIP:940308}} \\
		Known Uses            & \parbox[t]{0.9\columnwidth}{\cite{designing-service-applications}} \\
	\end{tabular}
	\label{tab:endpoint-patterns-1}
\end{table}

\begin{table}
	\caption{Timed Redelivery \index{Timed Redelivery|textbf} \label{ceip:redelivering_endpoint}}
	\begin{tabular}{l*{2}{l}r}
		\hline
		Pattern Name          & Timed Redelivery \index{Timed Redelivery|textbf} \\
		\hline
		Intent                & \parbox[t]{0.9\columnwidth}{Enable the endpoint/application to deliver a message asynchronously using redelivery instead of intermediate storage as guarantee.} \\
		Driving Question      & \parbox[t]{0.9\columnwidth}{How to ensure that a message will be received without intermediate storage, \eg in form of Redelivery on Exception \cite{DBLP:conf/edoc/RitterS14}} \\
		Context               & \parbox[t]{0.9\columnwidth}{Asynchronous communication with message delivery guarantees.} \\
		Solution              & \parbox[t]{0.9\columnwidth}{Instead of relying on intermediate storage and retry within the integration system, the application sends multiple instances of the same message with configurable timings until the actual receiver endpoint acknowledgements (\eg \emph{Quick Acknowledgement} \cite{DBLP:conf/dagstuhl/Hohpe06}) reach the sender. Requires an Idempotent \cite{Hohpe:2003:EIP:940308} and Commutative Receiver for certain message delivery semantics \cite{ritter2015integration}} \\
		Data Aspects          & \parbox[t]{0.9\columnwidth}{Timings, acknowledgements} \\
		Result                & \parbox[t]{0.9\columnwidth}{Send copies of the same message asynchronously until the receiver's acknowledgement is reaches the sender.} \\
		Related Patterns      & \parbox[t]{0.9\columnwidth}{Retry \cite{DBLP:conf/dagstuhl/Hohpe06}, Redelivery on Exception \cite{DBLP:conf/edoc/RitterS14}.} \\
		Known Uses            & \parbox[t]{0.9\columnwidth}{--} \\
	\end{tabular}
	\label{tab:endpoint-patterns-1}
\end{table}

\begin{figure}[tbh]
	\centering
	\includegraphics[width=0.9\columnwidth]{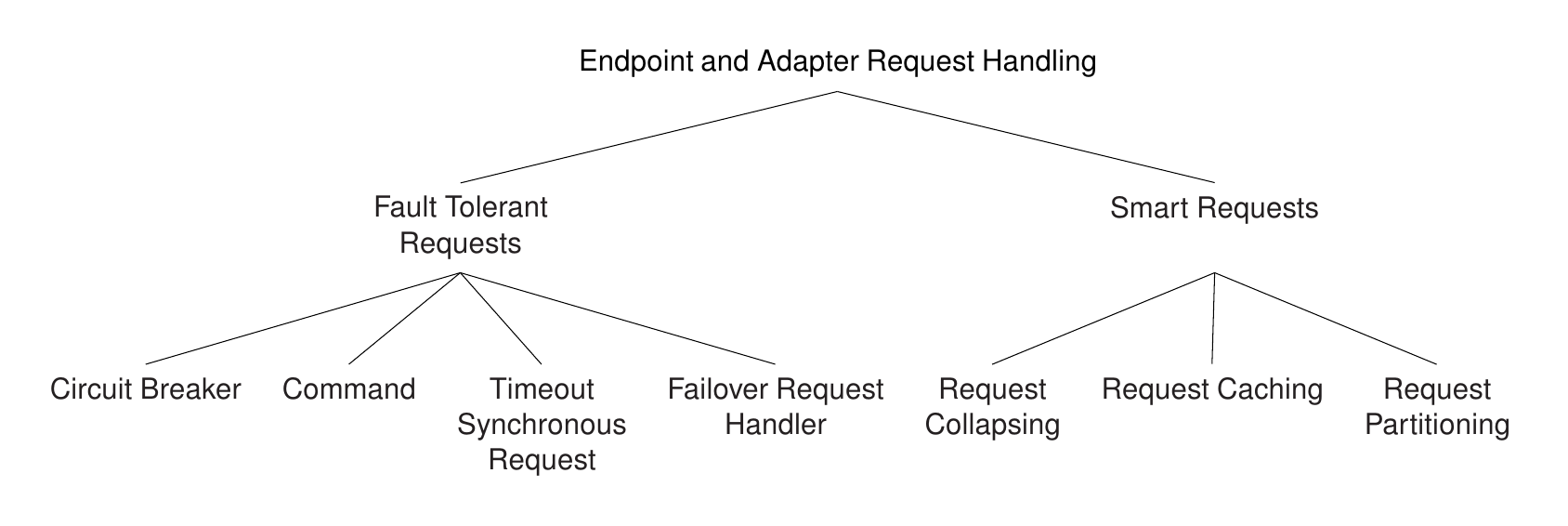}
	\caption{Integration Adapter and Endpoint Request Patterns.}
	\label{fig:requests-overview}
\end{figure}

Recent practical advances for cloud-based \emph{Microservice} architectures \cite{Newman15}, \eg in the area of fault-tolerance like \emph{Hystrix} \cite{hystrix}, can be mapped to integration systems and summarized to general patterns as shown in \cref{fig:requests-overview}.

Thereby the related aspects of fault-tolerance and smart requests are differentiated. For fault-tolerance, one wants to make request more tangible (cf. \cref{tab:endpoint-patterns-2}), to penalize long running requests (cf. \cref{tab:endpoint-patterns-1}), break them by timeouts (cf. \cref{tab:endpoint-patterns-3}) and find alternative endpoints that can answer the requests instead (cf. \cref{tab:endpoint-patterns-4}).

\begin{table}
	\caption{Circuit Breaker \index{Circuit Breaker|textbf} \label{ceip:circuit_breaker}}
	\begin{tabular}{l*{2}{l}r}
		\hline
		Pattern Name          & Circuit Breaker \index{Circuit Breaker|textbf} \\
		\hline
		Intent                & \parbox[t]{0.9\columnwidth}{Stop all requests to a particular endpoint/application for a period of time} \\
		Driving Question      & \parbox[t]{0.9\columnwidth}{How to stop all requests to a particular endpoint/application for a period of time?} \\
		Context               & \parbox[t]{0.9\columnwidth}{Communication with endpoints/applications} \\
		Solution              & \parbox[t]{0.9\columnwidth}{Provide capabilities to monitor the error percentage, when communicating with a particular endpoint and automatically or manually prevent its invocation for a period of time.} \\
		Data Aspects          & \parbox[t]{0.9\columnwidth}{Timing, monitoring statistics} \\
		Result                & \parbox[t]{0.9\columnwidth}{Reduces number of failed message exchanges.} \\
		Related Patterns      & \parbox[t]{0.9\columnwidth}{Electronic circuits} \\
		Known Uses            & \parbox[t]{0.9\columnwidth}{\enquote{Circuit Breaker} \cite{hystrix}, conceptually from \cite{Nygard07}} \\
	\end{tabular}
	\label{tab:endpoint-patterns-1}
\end{table}


\begin{table}
	\caption{Command \index{Command|textbf} \label{ceip:command}}
	\begin{tabular}{l*{2}{l}r}
		\hline
		Pattern Name          & Command \index{Command|textbf} \\
		\hline
		Intent                & \parbox[t]{0.9\columnwidth}{Make remote calls tangible.} \\
		Driving Question      & \parbox[t]{0.9\columnwidth}{How to make remote calls tangible?} \\
		Solution              & \parbox[t]{0.9\columnwidth}{Wrap remote calls into one command context.} \\
		Result                & \parbox[t]{0.9\columnwidth}{The command context allows to, \eg timeout and monitor a request.} \\
		Related Patterns      & \parbox[t]{0.9\columnwidth}{Similar to \enquote{Adapter Flow} in \cite{ritter2015integration}, Request Collapsing}\\
		Known Uses            & \parbox[t]{0.9\columnwidth}{\enquote{Reactive Command} \cite{hystrix}} \\
	\end{tabular}
	\label{tab:endpoint-patterns-2}
\end{table}


\begin{table}
	\caption{Timeout Synchronous Request \index{Timeout Synchronous Request|textbf} \label{ceip:timeout_request}}
	\begin{tabular}{l*{2}{l}r}
		\hline
		Pattern Name          & Timeout Synchronous Request \index{Timeout Synchronous Request|textbf} \\
		\hline
		Intent                & \parbox[t]{0.9\columnwidth}{Timeout remote calls that take longer than a configured threshold.} \\
		Driving Question      & \parbox[t]{0.9\columnwidth}{How to timeout remote calls that take longer than a configured threshold?} \\
		Context               & \parbox[t]{0.9\columnwidth}{Long blocking requests destabilize the scenario and potentially the integration system.} \\
		Solution              & \parbox[t]{0.9\columnwidth}{Apply a timeout to each synchronous request.} \\
		Data Aspects          & \parbox[t]{0.9\columnwidth}{Timing} \\
		Result                & \parbox[t]{0.9\columnwidth}{Unblocks the scenario and system.} \\
		Related Patterns      & \parbox[t]{0.9\columnwidth}{Command, \enquote{Delayed Redelivery}, \enquote{Stop Operation} both from \cite{DBLP:conf/edoc/RitterS14}, Message Expiration \cite{Hohpe:2003:EIP:940308}} \\
		Known Uses            & \parbox[t]{0.9\columnwidth}{\enquote{Timeout} \cite{hystrix}, conceptually from \cite{Nygard07}} \\
	\end{tabular}
	\label{tab:endpoint-patterns-3}
\end{table}



\begin{table}
	\caption{Failover Request Handler \index{Failover Request Handler|textbf} \label{ceip:failover_request}}
	\begin{tabular}{l*{2}{l}r}
		\hline
		Pattern Name          & Failover Request Handler \index{Failover Request Handler|textbf} \\
		\hline
		Intent                & \parbox[t]{0.9\columnwidth}{Constructively handling failed requests.} \\
		Driving Question      & \parbox[t]{0.9\columnwidth}{How to handle failed or timed-out requests or short-circuits constructively?} \\
		Solution              & \parbox[t]{0.9\columnwidth}{Perform fallback logic when a request fails, is rejected, times-out or short-circuits.} \\
		Data Aspects          & \parbox[t]{0.9\columnwidth}{Configuration of failover channels or endpoints/adapters} \\
		Variations            & \parbox[t]{0.9\columnwidth}{Applicable to endpoints and adapters.} \\
		Example               & \parbox[t]{0.9\columnwidth}{If there is a communication error with the currently selected endpoint, then the request handler automatically fails-over to the next endpoint in the list.} \\
		Related Patterns      & \parbox[t]{0.9\columnwidth}{Command, Failover Router \cite{DBLP:conf/edoc/RitterS14}} \\
		Known Uses            & \parbox[t]{0.9\columnwidth}{\enquote{Failover Client} \cite{apache-flume}} \\
	\end{tabular}
	\label{tab:endpoint-patterns-4}
\end{table}



The smart handling of requests helps to stabilize the system as well. However, the main focus lies on an intelligent way of requesting information from endpoints, \eg by combining several (cross-scenario) requests to one (cf. \cref{tab:endpoint-patterns-5}), or leveraging already recently requested information for multiple subsequent requests (cf. \cref{tab:endpoint-patterns-6}) or separating request aspects to multiple requesters (cf. \cref{tab:endpoint-patterns-7}).

\begin{table}
	\caption{Request Collapsing \index{Request Collapsing|textbf} \label{ceip:request_collapsing}}
	\begin{tabular}{l*{2}{l}r}
		\hline
		Pattern Name          & Request Collapsing \index{Request Collapsing|textbf} \\
		\hline
		Intent                & \parbox[t]{0.9\columnwidth}{Reduce the number of endpoint calls.} \\
		Driving Question      & \parbox[t]{0.9\columnwidth}{How to reduce the number of endpoint calls?} \\
		Solution              & \parbox[t]{0.9\columnwidth}{Collapse multiple requests into a single endpoint dependency call.} \\
		Data Aspects          & \parbox[t]{0.9\columnwidth}{Data aggregation, selection and projection handling} \\
		Variations            & \parbox[t]{0.9\columnwidth}{Applicable to endpoints and adapters.} \\
		Related Patterns      & \parbox[t]{0.9\columnwidth}{Command, similar to \enquote{Adapter Flow} in \cite{ritter2015integration}} \\
		Known Uses            & \parbox[t]{0.9\columnwidth}{\enquote{Request Collapsing} \cite{hystrix}, conceptually supported by \cite{Vrhovnik11} for database access from business process engines.} \\
	\end{tabular}
	\label{tab:endpoint-patterns-5}
\end{table}

\begin{table}
	\caption{Request Caching \index{Request Caching|textbf} \label{ceip:request_caching}}
	\begin{tabular}{l*{2}{l}r}
		\hline
		Pattern Name          & Request Caching \index{Request Caching|textbf} \\
		\hline
		Intent                & \parbox[t]{0.9\columnwidth}{Reduce the amount of duplicate requests to the same endpoint.} \\
		Driving Question      & \parbox[t]{0.9\columnwidth}{How to reduce the amount of duplicate requests (from several scenarios) to the same endpoint?} \\
		Context               & \parbox[t]{0.9\columnwidth}{Several integration scenarios might request the same information from the same endpoint (concurrently).} \\
		Solution              & \parbox[t]{0.9\columnwidth}{Cache request contexts/keys to connect only once and feed many integration scenarios with the required information.} \\
		Data Aspects          & \parbox[t]{0.9\columnwidth}{Data aggregation, selection and projection handling, (persistent) caching, request correlation} \\
		Variations            & \parbox[t]{0.9\columnwidth}{Applicable to endpoints and adapters.} \\
		Related Patterns      & \parbox[t]{0.9\columnwidth}{Command, Data Store} \\
		Known Uses            & \parbox[t]{0.9\columnwidth}{similar to \enquote{Request Caching} in \cite{hystrix}, conceptually supported by \cite{Vrhovnik11} for database access from business process engines.} \\
	\end{tabular}
	\label{tab:endpoint-patterns-6}
\end{table}


\begin{table}
	\caption{Request Partitioning \index{Request Partitioning|textbf} \label{ceip:request_partitioning}}
	\begin{tabular}{l*{2}{l}r}
		\hline
		Pattern Name          & Request Partitioning \index{Request Partitioning|textbf} \\
		\hline
		Intent                & \parbox[t]{0.9\columnwidth}{Isolate request dependencies and limit concurrent access to them.} \\
		Driving Question      & \parbox[t]{0.9\columnwidth}{How to isolate requests and limit concurrent access to any of them?} \\
		Solution              & \parbox[t]{0.9\columnwidth}{Manually or automatically isolate request dependencies.} \\
		Data Aspects          & \parbox[t]{0.9\columnwidth}{Split queries (where possible) and merge results} \\
		Result                & \parbox[t]{0.9\columnwidth}{Partitions request failures as well for a more robust system: by partitioning requests to endpoints, errors are confined to one request aspect (\eg get business partner header data) as opposed to cancelling the entire request (\eg business partner header and associated documents).} \\
		Variations            & \parbox[t]{0.9\columnwidth}{The partitions can be hardware redundancy, binding certain processes to certain CPUs, segmenting different areas of business functionality to different server farms, or partitioning requests into different information aspect groups for different functionality.} \\
		Related Patterns      & \parbox[t]{0.9\columnwidth}{Bulkhead, Command, Request Collapsing} \\
		Known Uses            & \parbox[t]{0.9\columnwidth}{\enquote{Isolation} in \cite{hystrix}, conceptually similar to \enquote{Bulkhead} from \cite{Nygard07}} \\
	\end{tabular}
	\label{tab:endpoint-patterns-7}
\end{table}

\newpage
\subsection{Composition}
\label{sec:composition}

The composition of patterns to integration processes requires hierarchical partitioning for reuse and scoping (cf. \cref{tab:composition-1}), as well as the creation of process variants (cf. \cref{tab:composition-2}).

\begin{table}
	\caption{Integration Subprocess \index{Integration Subprocess|textbf} \label{ceip:subprocess}}
	\begin{tabular}{l*{2}{l}r}
		\hline
		Pattern Name          & Integration Subprocess \index{Integration Subprocess|textbf} \\
		\hline
		Intent                & \parbox[t]{0.9\columnwidth}{Reuse, scope integration process.} \\
		Driving Question      & \parbox[t]{0.9\columnwidth}{How to reuse, scope an integration process?} \\
		Solution              & \parbox[t]{0.9\columnwidth}{Hierarchy of processes. A sub-process is attached to one parent process (\ie integration process, sub-process)} \\
		Data Aspects          & \parbox[t]{0.9\columnwidth}{Process context} \\
		Result                & \parbox[t]{0.9\columnwidth}{Integration processes can be partitioned into smaller processes for re-use or different scoping.} \\
		Variations            & \parbox[t]{0.9\columnwidth}{Shared sub-process that allows for reusing an Integration Subprocess by other integration processes.} \\
		Related Patterns      & \parbox[t]{0.9\columnwidth}{Exception Subprocess} \\
		Known Uses            & \parbox[t]{0.9\columnwidth}{\eg \cite{ibm,microsoft,sap-hci}} \\
	\end{tabular}
	\label{tab:composition-1}
\end{table}

\begin{table}
	\caption{Integration Process Template \index{Integration Process Template|textbf} \label{ceip:template}}
	\begin{tabular}{l*{2}{l}r}
		\hline
		Pattern Name          & Integration Process Template \index{Integration Process Template|textbf} \\
		\hline
		Intent                & \parbox[t]{0.9\columnwidth}{Create variants of integration processes.} \\
		Driving Question      & \parbox[t]{0.9\columnwidth}{How to create variants of an integration process?} \\
		Solution              & \parbox[t]{0.9\columnwidth}{Create structural representation of an integration process and externalize all configurations. The structure and configurations can be changed to create a new variant or instance of the process.} \\
		Data Aspects          & \parbox[t]{0.9\columnwidth}{Process structure, configuration parameters} \\
		Result                & \parbox[t]{0.9\columnwidth}{The structural aspects are separated from the instance configuration.} \\
		Related Patterns      & \parbox[t]{0.9\columnwidth}{-} \\
		Known Uses            & \parbox[t]{0.9\columnwidth}{Template Integration Process \cite{ibm}, Snapshot \cite{jitterbit}, Blueprint \cite{cloudpipes}} \\
	\end{tabular}
	\label{tab:composition-2}
\end{table}

\newpage
\section{SAP HANA Cloud Integration eDocument}
\label{sec:examples}

\begin{figure}[ht!]
	\centering
	\includegraphics[width=0.7\columnwidth]{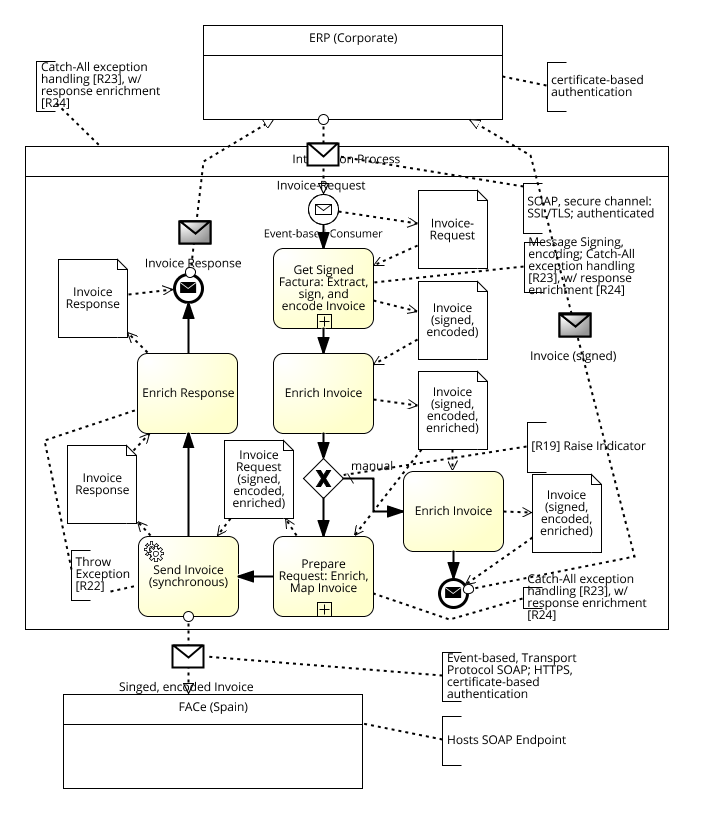}
	\caption{eDocument Send Invoice (Spain), in BPMN using Signavio.}
	\label{fig:edocument_spain_scenario} 
\end{figure}

The SAP HCI eDocument solution allows to create, process and manage electronic documents for country-specific requirements.
\Cref{fig:edocument_spain_scenario} shows an eDocument scenario for corporates sending their factura invoice to Spanish eDocument authorities FACe (\emph{Factura-e})\footnote{Spain \emph{Factura-e}: \url{http://www.facturae.gob.es/}} using SAP HCI.
Hereby, the electronic invoice is a statement with the same legal effect as a paper invoice, however, transmitted electronically.
The corporates have to send the invoice whose format is specified by the Spanish authorities to the FACe by 15th of January each year.
The corporates send a SOAP message, containing the factura to HCI, where the factura is extracted, signed (\ie SHA1/RSA), and encoded according to the regulations (\ie Base64), \eg within a BPMN \emph{SubProcess}.
The message contains routing information about the manual or automatic delivery to the FACe.
In case of \enquote{manually} delivery an indication is raised in form of an alert into a monitor for manual post-processing. Otherwise, the message is converted into the required SOAP request format and sent to the FACe. Thereafter, a response to the corporate is prepared and sent. In case of an exceptional situation during processing (\ie throw exception \cite{DBLP:conf/edoc/RitterS14}), all exceptions are caught (\ie catch-all \cite{DBLP:conf/edoc/RitterS14})
and processed in an exception flow (\ie exception processing \cite{DBLP:conf/edoc/RitterS14}). The signing, indication raising, and exception handling are not in EIP.

\begin{figure}[ht!]
	\begin{center}
		\label{fig:topic1}
		\includegraphics[width=0.7\textwidth]{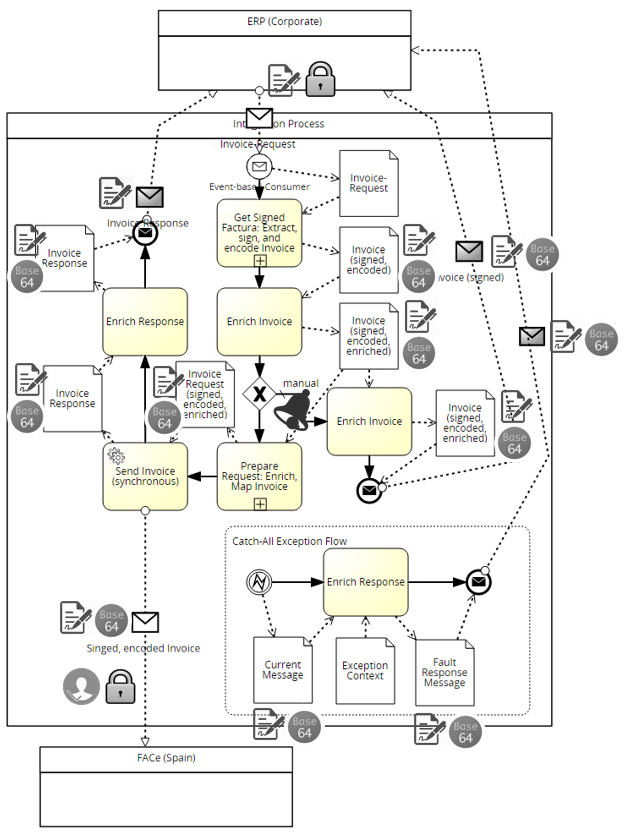}
	\end{center}\par\medskip
	\footnotesize
	\textbf{Used Annotations:} \includegraphics[width=0.03\textwidth]{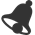} raise indicator,
	\includegraphics[width=0.025\textwidth]{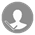} authenticated,
	\includegraphics[width=0.025\textwidth]{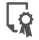} certificate-based,
	\includegraphics[width=0.025\textwidth]{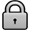} encrypted,
	\includegraphics[width=0.025\textwidth]{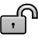} decrypted,
	\includegraphics[width=0.025\textwidth]{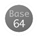} encoded,
	\includegraphics[width=0.025\textwidth]{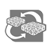} redundant,
	\includegraphics[width=0.025\textwidth]{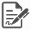} signed,
	\includegraphics[width=0.025\textwidth]{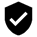} verified.
	\caption{CIP Application (BPMN + selected CIP icons + annotations)}
	\label{fig:topics}
\end{figure}

\Cref{fig:topics} shows the synthesis of the eDocuments scenario with the new patterns partially as imperative BPMN syntax combined with a declarative, icon notation. The icons are BPMN-conpliant and can be realized as configuration tables as part of the used BPMN elements.

\bibliographystyle{abbrv}
\bibliography{pattern-supplement}

\newpage
\renewcommand{\indexname}{List of Patterns}
\printindex

\newpage
\appendix

\section{Pattern Format}
In this section the pattern format used in this paper is described. The format,  shown in \cref{tab:template}, is defined similar to existing pattern formats of the EIP \cite{Hohpe:2003:EIP:940308} and related pattern descriptions (\eg from cloud computing \cite{DBLP:books/sp/FehlingLRSA14}).

\begin{table}
	\caption{Pattern Format Template \index{Pattern Format Template|textbf} \label{beip:template}}
	\begin{tabular}{l*{2}{l}r}
		\hline
		Pattern Name          & Name \label{beip:encrypted-message} \\
		\hline
		Intent                & \parbox[t]{0.9\columnwidth}{at the beginning of each pattern, its purpose and goal is shortly stated, to describe what the solution represented by the pattern contains.} \\
		Driving Question      & \parbox[t]{0.9\columnwidth}{This question captures the problem that is answered by the pattern. Stating this question at the beginning allows readers to identify if the pattern fits the problem they have in a concrete use case.} \\
		Context (optional)    & \parbox[t]{0.9\columnwidth}{This section describes the environment and forces leading to the problem solved by the pattern. It also may describe why naive solutions can be unsuccessful or suboptimal. Other patterns may be referenced here.} \\
		Solution              & \parbox[t]{0.9\columnwidth}{The solution section briefly states how the pattern solves the problem raised by the driving question. It is kept brief, because readers shall be enabled to quickly read the intent, question, and solution sections to get an idea what the pattern is doing in detail. The solution section is commonly closed with a sketch depicting the architecture of the solution.} \\
		Data Aspects (optional) &  \parbox[t]{0.9\columnwidth}{All message, pattern content and configuration aspects are described in this section.}\\
		Result (optional)     & \parbox[t]{0.9\columnwidth}{In this section, the solution is elaborated in greater detail. The architecture proposed by the sketch is described and the behavior of the application after implementation of the pattern is discussed. New challenges that may arise after a pattern has been applied may also be included here, together with references to other patterns addressing these new challenges.} \\
		Variations (optional) & \parbox[t]{0.9\columnwidth}{Often, patterns can be applied in slightly different forms. If the differences of these variations are not significant enough to justify their description in a separate pattern, they are covered in the variations section.} \\
		Example (optional)    & \parbox[t]{0.9\columnwidth}{This section gives an illustrative example of the pattern.} \\
		Related Patterns      & \parbox[t]{0.9\columnwidth}{Several patterns are often applied together as they are solving related problems, but the application of one pattern may also exclude other patterns from being applicable.	These interrelations of patterns are described in this section. It, therefore, forms the structure of the integration pattern language and guides readers through the set of patterns.} \\
		Known Uses            & \parbox[t]{0.9\columnwidth}{Existing applications implementing a pattern, products offering a pattern or supporting its implementation, are covered here exemplarily.} \\
	\end{tabular}
	\label{tab:template}
\end{table}

\newpage

\end{document}